\documentclass[12pt,leqno]{article}

\usepackage{graphicx}
\usepackage{epsfig} 
\usepackage{amsmath, amsfonts,  float}
\usepackage{indentfirst}
\usepackage{ifpdf}  
\usepackage{epstopdf}
\usepackage{color}
\usepackage{amsthm}
\usepackage{latexsym}
\usepackage{multimedia}
\usepackage{epstopdf}

\makeatletter
\renewcommand\paragraph{%
  \@startsection{paragraph}{4}{0mm}%
   {-\baselineskip}%
   {.5\baselineskip}%
   {\normalfont\normalsize\bfseries}}
\makeatother

\begin{document}

\theoremstyle{remark}
\newtheorem{remark}{Remark}[section]
%
%

\newcommand{\bn}{{\bf n}}
\newcommand{\bnabla}{{\boldsymbol{\nabla}}}
\newcommand{\bu}{{\bf u}}
\newcommand{\bU}{{\bf U}}
\newcommand{\bPhi}{{\bf \Phi}}
\newcommand{\bPsi}{{\bf \Psi}}
\newcommand{\bv}{{\bf v}}
\newcommand{\bw}{{\bf w}}
\newcommand{\bx}{{\bf x}}
\newcommand{\bV}{{\bf V}}
\newcommand{\bB}{{\bf B}}
\newcommand{\bz}{{\bf 0}}
\newcommand{\cth}{{\mathcal{T}_h}}
\newcommand{\ct}{{\mathcal{T}}}
\newcommand{\f}{{\bf f}}
\newcommand{\g}{{\bf g}}
\newcommand{\bgamma}{{\boldsymbol{\gamma}}}
\newcommand{\blambda}{{\boldsymbol{\lambda}}}
\newcommand{\gx}{{\bgamma_{-} (0, \Delta t)}}
\newcommand{\into}{{\displaystyle{\int_{\Omega}}}}
\newcommand{\intG}{{\displaystyle{\int_{\bgamma}}}}
\newcommand{\oo}{{\overline{\Omega}}}

\newcommand{\ox}{{\omega \times (0, \Delta t)}}
\newcommand{\oxot}{{\omega \times (0,T)}}

\newcommand{\Ox}{{\Omega \times (0, \Delta t)}}
\newcommand{\Oxot}{{\Omega \times (0,T)}}
\newcommand{\R}{{I\!\!R}}
\renewcommand{\arraystretch}{1.4}

\noindent{\Large \bf  Numerical study of transitions in lid-driven  flows in semicircular cavities}

\vskip 2ex
\noindent{\normalsize  Tsorng-Whay Pan$^{a,}$\footnote{Corresponding author: e-mail:  
		tpan@uh.edu}, Ang Li$^b$, and  Shang-Huan Chiu$^c$}
\vskip 1ex
\noindent {\small $^a$ Department of Mathematics, University of Houston, Houston, Texas 77204, USA} 
\vskip 1ex
\noindent {\small $^b$ Department of Mathematics, Lane College, Jackson, TN 38301, USA}
\vskip 1ex
\noindent {\small $^c$ Department of Mathematics, Lehigh University, Bethlehem, PA, 18015, USA}



\vskip 4ex
\noindent {\bf Abstract} \ 
In this article, three-dimensional (3D) lid-driven flows in  semicircular cavities are studied.  
The numerical solution of the Navier-Stokes equations  modeling incompressible viscous 
fluid flow in  cavities is obtained  via a methodology combining a first-order accurate
operator-splitting scheme, a fictitious domain formulation, and finite element space approximations.   
The critical Reynolds numbers ($Re_{cr}$) for having oscillatory  flow (a Hopf bifurcation)
are obtained. The associated oscillating motion in a semicircular cavity with length equal to width 
has been studied in detail.  Based on the averaged velocity field in one period of oscillating motion,
the flow difference (called oscillation mode) between the velocity field and averaged one at several time instances 
in such period shows almost the same flow pattern for the Reynolds numbers close to $Re_{cr}$. This
oscillation mode in a semicircular cavity shows a close similarity to the one obtained in a 
shallow cavity, but with some difference in a shallow cavity which is triggered by the presence 
of two vertical side walls and downstream wall.

\vskip 4ex



\baselineskip 14pt

\setlength{\parindent}{1.5em}

\section{Introduction}

\normalsize
Lid-driven cavity flow is a classical benchmark flow problem  for validating numerical methods 
and comparing results obtained from laboratory and computational experiments due to its 
geometrical simplicity and unambiguous boundary conditions (e.g., see \cite{Shankar2000}, 
\cite{Guermond2002}, and \cite{Kuhlmann2019}). This flow problem is important to the basic  
study of fluid mechanics, including boundary layers, primary vortex, secondary flows 
(such as the corner vortices and Taylor-Goertler-like vortices),  various instabilities 
and transitions, and turbulence;  this flow system is also relevant to many industrial 
applications (e.g., see \cite{Shankar2000} and \cite{Aidun1991}).  
Most lid-driven cavity flows were studied in two-dimensional (2D) and three-dimensional (3D) 
rectangular cavities; but, e.g.,  some other shapes like those with triangular, polar or 
sectorial cross section were also considered  (see a review article \cite{Kuhlmann2019} 
for some of those non-rectangular cases).  
For cavities with a circular shape boundary, there are fewer results available (compared to 
rectangular cavities) as pointed out in \cite{Kuhlmann2019} and  \cite{Gonzalez2017}. 
In \cite{Gerrits1996} and \cite{GlowPan2022}, lid-driven cavity flows have been 
obtained in a hemispherical cavity and its transition has been studied in \cite{GlowPan2022}. 
In \cite{Belhachmi2004}, a spectral element discretization was developed to solve lid-driven 
flows in a full disk with  a moving circular lid. A similar one was also considered later in  
\cite{Gonzalez2017}, and  a circular cavity with an horizontal top boundary was also studied 
in \cite{Gonzalez2017}. Although two-dimensional flows were obtained in \cite{Gonzalez2017},
their 3D linear stability analyses were done and  confirmed by
spectral direct numerical simulations with periodic flows in the spanwise direction.
In a 2D semicircular cavity,  a classic finite element approach was used to solve 
lid-driven flows in \cite{Glow2006}. Such method has shown no difficulty at capturing
the formation of primary, secondary and tertiary vortices as Re increases;
it also has the capability in capturing the transition from steady flow to 
oscillatory flow (a Hopf bifurcation phenomenon).    In \cite{Migeon2000},
a 3D semicircular cavity was one of several cavity shapes used to study experimentally 
the shape influence on the birth and evolution of recirculating flow 
structures in cavities. The evolution of lid-driven flow in a semicircular cavity
was studied up to the dimensionless time $t$=12 since semicircular flow seems to 
reach its steady state as claimed in  \cite{Migeon2000} and flow transition to 
oscillatory one was not mentioned.

It is known that,  depending on the solution method, boundary conditions and mesh size used 
in simulation, the critical Reynolds number (Re$_{cr}$) for the occurrence of transition from  
steady flow to  oscillatory flow  varies in cavities.
For example,   Iwatsu {\it et al.}  \cite{Iwatsu1990}  obtained 
numerically a pair of Taylor-G\"ortler-like  vortices for  a cubic lid-driven cavity flow 
at Re=2000. Giannetti {\it et al.} \cite{Giannetti2009} also obtained that the cubic 
lid-driven cavity flow  becomes unstable for  Re just above 2000 via the three-dimensional 
global linear stability analysis. Feldman and Gelfgat \cite{Feldman2010} 
obtained the critical Reynolds number for  transition occurring at Re$_{cr}=1914$.  
Liberzon {\it et al.} \cite{Liberzon2011} experimentally obtained the critical Reynolds 
number is  in the range $[1700, 1970]$.   In \cite{PanChiuGuoHe2023}, 
Pan {\it et al.}  found the critical Reynolds number value is between 1894 and 1895 
for lid-driven flows in a cube.  In this article, we have studied numerically the transition 
from steady flow to oscillatory one  in  semicircular cavities. We have applied  
a first-order accurate operator-splitting scheme, the Lie scheme (e.g., see \cite{GlowPan2022} and
\cite{OS2016} for its details), with a fictitious domain approach and finite element method to  obtain
numerical solutions of the Navier-Stokes equations. This numerical methodology is an extension 
of the investigations reported in  \cite{Pan2008} and \cite{Pan2000}. The resulting methodology is 
easy to implement and quite modular since, at each time step,  one has to solve a sequence 
of four simpler sub-problems.  To investigate the mode associated with the transition from 
steady flow to oscillatory flow,  we have focused on the flow  fields at Reynolds numbers  close to Re$_{cr}$. 
The difference of  flow field  with respect to the averaged flow field in one period of the 
oscillation has been studied and compared with those that occurred in a shallow cavity with 
unit square base.  The outline of this paper is as follows:  We first introduce the formulation 
of the flow problem  and then the numerical method briefly in Section 2. In Section 3, numerical 
results obtained for  lid-driven flows in semicircular cavities are presented. Then   the 
transition from steady flow to oscillatory flow has been studied, especially on the comparison of 
oscillatory  modes in two different cavities.  Conclusions are summarized in Section 4.

\section{Problem formulation}
\begin{figure} [ht]
\begin{center}
\leavevmode
\includegraphics[width=2.in]{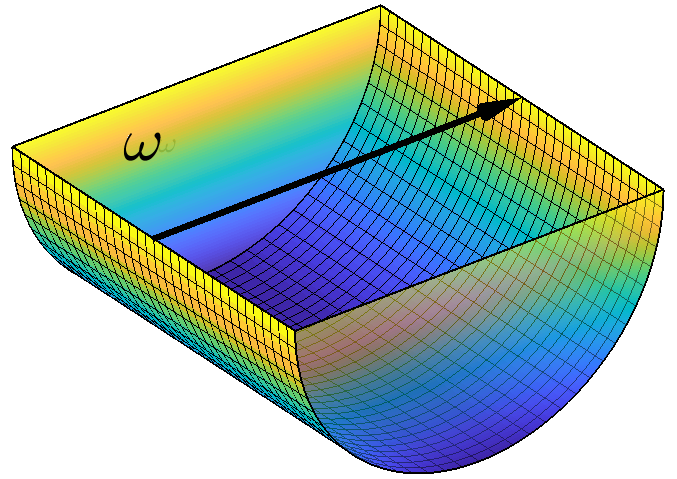} 
\end{center}
\caption{An example of a semicircular cavity $\omega$.}\label{fig.1}
\end{figure}
The governing equations for modeling incompressible viscous Newtonian fluid flow in a cavity 
$\omega \subset \R^3$ (see Figure \ref{fig.1}) for $T>0$ are the Navier-Stokes equations, namely
\begin{eqnarray}
&\dfrac{\partial \bu}{\partial t} - \nu \Delta \bu + (\bu \cdot
\bnabla) \bu + \bnabla p = {\bf 0} \ \ in \ \ \oxot,  \label{eqn:1.1}\\
&\bnabla \cdot \bu = 0 \ \ in \ \ \oxot,\label{eqn:1.2}\\
&\bu(0) = \bu_0, \ \ with \ \ \bnabla \cdot \bu_0 = 0,\label{eqn:1.3}\\
&\bu = \bu_B(\bx) \ \ on \ \ \partial\oxot \ \ with \ \ \displaystyle\int_{\partial\omega} \bu_B 
\cdot \bn\, d\bgamma = 0 \ \ on \ \ (0,T),\label{eqn:1.4}
\end{eqnarray}
where $\bu$ and $p$ are the flow velocity and pressure, 
respectively, $\nu$ is a viscosity coefficient,  
and $\bn$ is the unit outward normal vector at the boundary $\bgamma=\partial \omega$.  
For lid-driven flows considered in this article, the boundary data $\bu_B(\bx)$ is $(1,0,0)^t$ 
on the top moving lid  and zero elsewhere on the boundary of $\omega$. We denote by $v(t)$ the 
function $\bx \to v(\bx,t)$, $\bx$ being the generic point of $\R^3$.

\begin{figure} [ht]
\begin{center}
\leavevmode
\includegraphics[width=2.in]{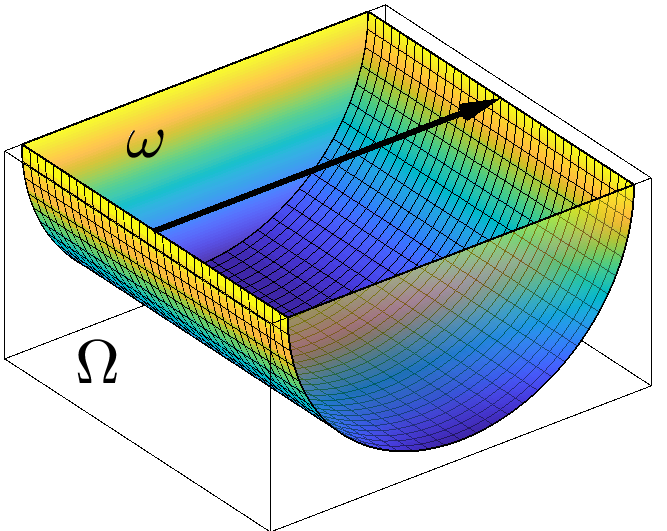} 
\end{center}
\caption{An example of a semicircular cavity $\omega$   embedded into a simple shape
fictitious domain $\Omega$.}\label{fig.2}
\end{figure}

To solve problem (\ref{eqn:1.1})-(\ref{eqn:1.4}) numerically, we have first embedded the fluid 
flow domain $\omega$ into a larger simple shape domain $\Omega$ (so-called fictitious domain, 
see Figure \ref{fig.2}) and obtained its fictitious domain formulation 
\begin{eqnarray}
&\dfrac{\partial \bU}{\partial t} - \nu \Delta \bU + (\bU \cdot
\bnabla) \bU + \bnabla P = {\blambda } \ \ in \ \ \Oxot,  \label{eqn:1.1a}\\
&\bnabla \cdot \bU = 0 \ \ in \ \ \Oxot,\label{eqn:1.2a}\\
&\bU(0) = \bU_0, \ \ with \ \ \bnabla \cdot \bU_0 = 0,\label{eqn:1.3a}\\
&\bU = \bU_B(\bx) \ \ on \ \ \partial\Oxot \ \ with \ \ \displaystyle\int_{\partial\Omega} \bU_B 
\cdot \bn\, d\bgamma = 0 \ \ on \ \ (0,T),\label{eqn:1.4a}\\
&\bU={\bf 0}\ \  in \ \ \Omega\setminus{\bar \omega} \times (0, T),\label{eqn:1.5a}
\end{eqnarray}
where $\blambda$ is a distributed Lagrange multiplier, which vanish in $\omega$, and acts as a pseudo 
body force so that $\bU={\bf 0}$ is enforced in $\Omega\setminus{\bar \omega}$, $\bU_0|_{\omega}=\bu_0$, 
and$\bU_B$ is $(1,0,0)^t$ on the top moving lid of $\omega$ and zero elsewhere on the boundary of $\Omega$.
Actually, the above distributed Lagrange multiplier approach has been successfully applied to simulate 
the motion of particles freely moving in a fluid (see, e.g., \cite{RG1999}, \cite{RG2001}, \cite{Pan2023}).
Then via the   Lie scheme (see, e.g., see    \cite{GlowPan2022} and \cite{OS2016} for the details)
to obtain the numerical solution of  lid-driven flow problem, 
we have  time-discretized problem (\ref{eqn:1.1a})-(\ref{eqn:1.5a}) into a sequence of four sub-problems
for each time step, namely: (i) using a $L^2$-projection  Stokes solver \`a la Uzawa to force the incompressibility 
condition, (ii) an advection step, (iii) a diffusion step, and (iv) enforcement zero velocity outside
the cavity $\omega$. A similar one for simulating lid-driven flow in a hemispherical cavity can be found
in  \cite{GlowPan2022} (Ch. 7).  Lie scheme is first-order accurate in time, but
its low order time accuracy  is compensated by its modularity,  easy implementation,   
stability, and robustness properties. The first three steps were used to obtain numerical results 
of lid-driven flow in shallow cavities reported in \cite{PanChiuGuoHe2023}.

For the space discretization, we have used, as  in   \cite{glowinski2003} (Chapter 5) 
and \cite{bristeau1987}, a $P_1$-$iso$-$P_2$ (resp., $P_1$) finite element  approximation for
the velocity field (resp., pressure) defined on uniform ‘‘tetrahedral’’  meshes $\cth$ (resp., $\ct_{2h}$)
due to the simple shape of fictitious domain.
The resulting four sub-problems via the Lie scheme are very classical problems 
and each one of them can be solved by a variety of existing methods, this being one of the key points of the 
operator-splitting methodology. 
For the first one, an $L^2$-projection    (equivalent to a saddle-point problem),
it can be solved by an Uzawa/preconditioned conjugate gradient algorithm as discussed 
in \cite{glowinski2003} (Section 21).  
The advection problem at the second step is solved by a wave-like equation method 
(see, e.g.,   \cite{dean1997}, \cite{dean1998}, and  \cite{GlowPan2022} (Ch. 3)) which is explicit 
and does not introduce numerical dissipation. Since the advection problem is decoupled from the others, 
a sub-time step satisfying the CFL condition  can be chosen easily. A classical  elliptic problem 
at the third step can be solved easily. The last one is also a saddle-point problem, which is
solved by a conjugate gradient algorithm as discussed in, e.g., \cite{GlowPan2022} (Ch. 7).

\section{Numerical Results and discussions}


\begin{figure} [!t]
\begin{center}
\leavevmode
\includegraphics[width=2.3in]{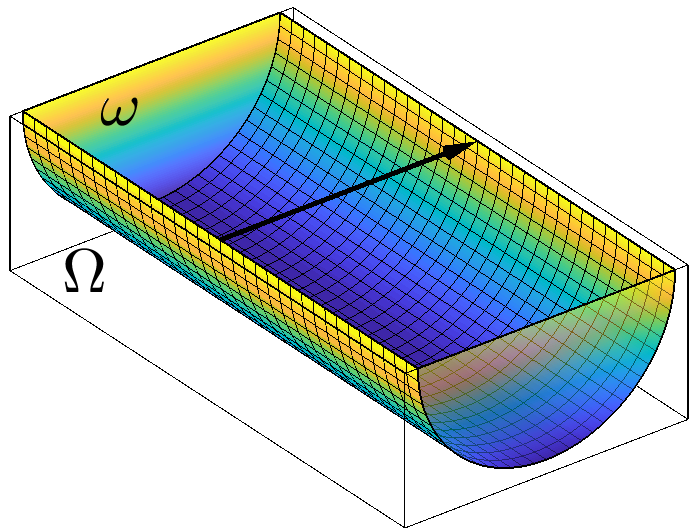} 
\end{center}
\caption{A semicircular cavity $\omega$ of width 2, depth 1, and height 0.5 (embedded into a fictitious domain $\Omega$) 
where the radius of circular sector is 0.5.}\label{fig.3}
\end{figure}
\vskip 4ex

\begin{figure} [!t]
\begin{center}
\leavevmode
\hskip -10pt \includegraphics[width=0.4in]{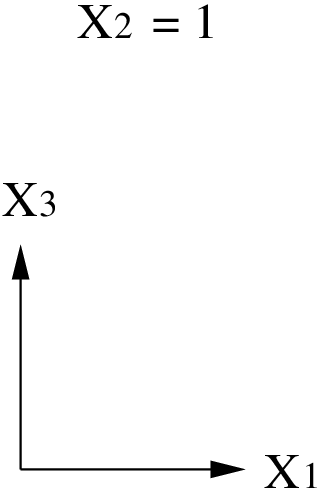}\
\includegraphics[width=2.2in]{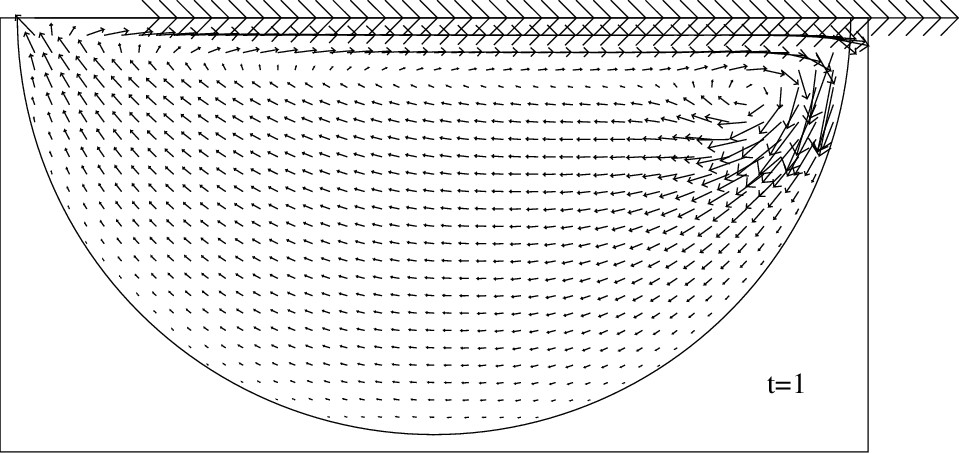} \  \includegraphics[width=2.2in]{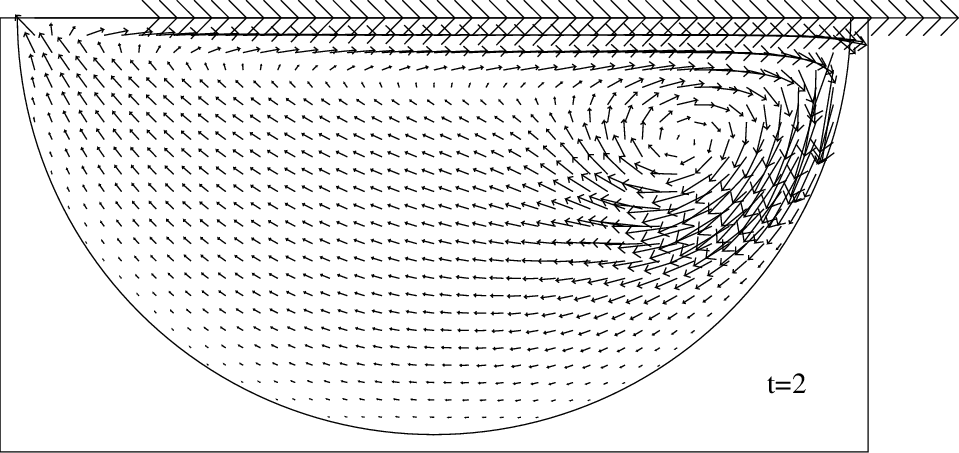} \\
\hskip 22pt \includegraphics[width=2.2in]{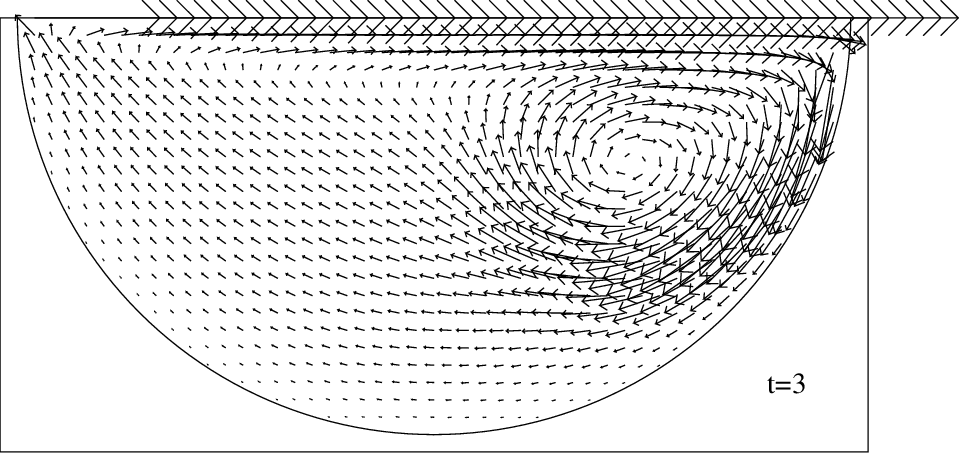} \  \includegraphics[width=2.2in]{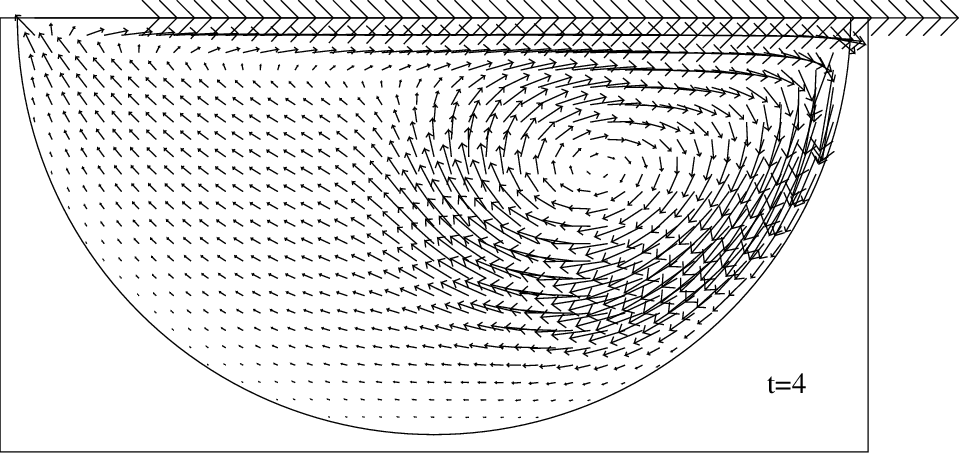} \\
\hskip 22pt \includegraphics[width=2.2in]{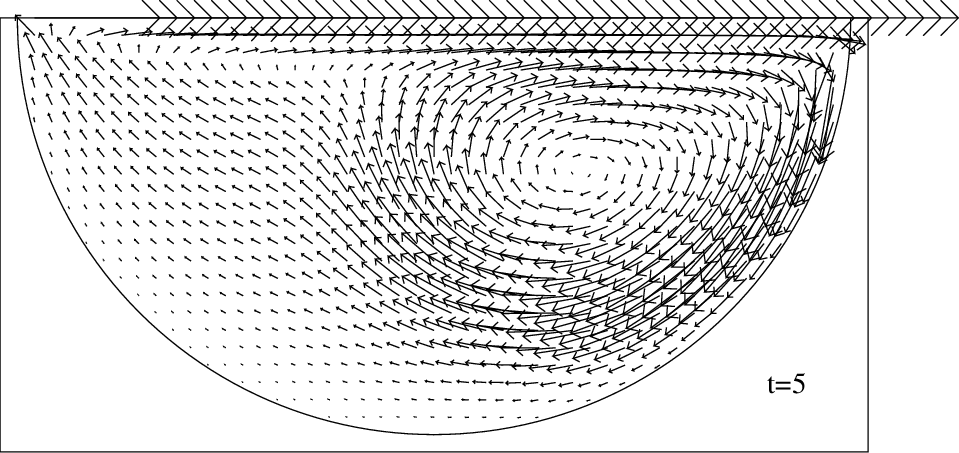} \  \includegraphics[width=2.2in]{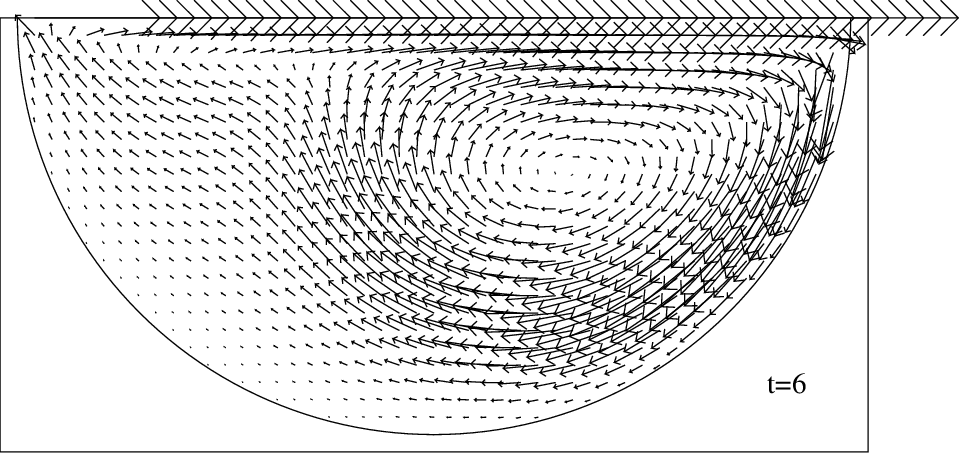} \\
\hskip 22pt \includegraphics[width=2.2in]{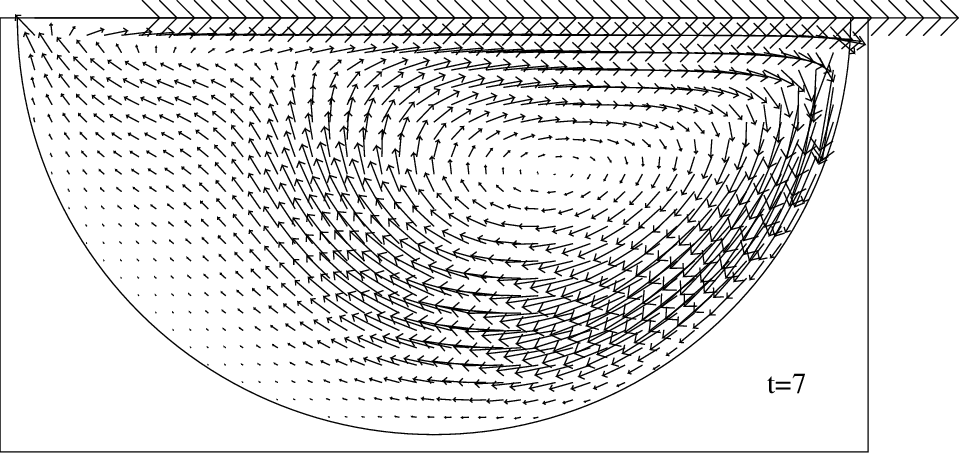} \  \includegraphics[width=2.2in]{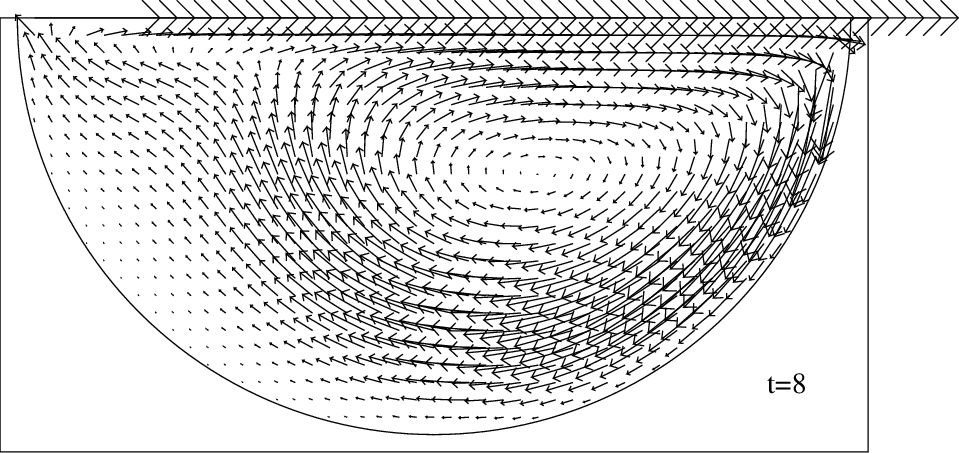} 
\end{center}
\caption{Projected velocity  vectors  on the plane $x_2=1$ at different times  for Re=1000.}\label{fig.4}
\end{figure}
\begin{figure}[t!]
\begin{center}
\leavevmode
\includegraphics[width=2.8in]{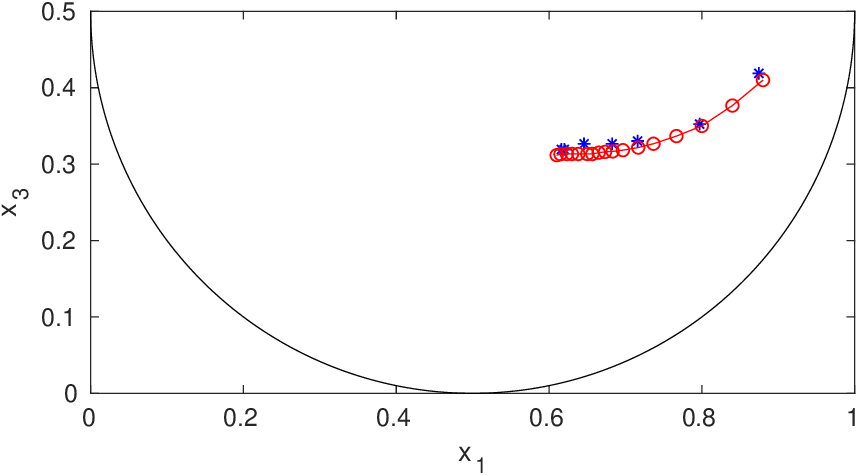}  
\end{center}
\caption{Primary   vortex-core trajectory in a semicircular cavity for Re=1000: 
Experimental data ($*$) taken from  \cite{Migeon2000} and  numerical results for 
$ 1 \le t \le 9$ sec (red solid line with $\circ$'s). }\label{fig.5}
\end{figure}

\begin{figure} [!t]
\begin{center}
\leavevmode
\includegraphics[width=1.8in]{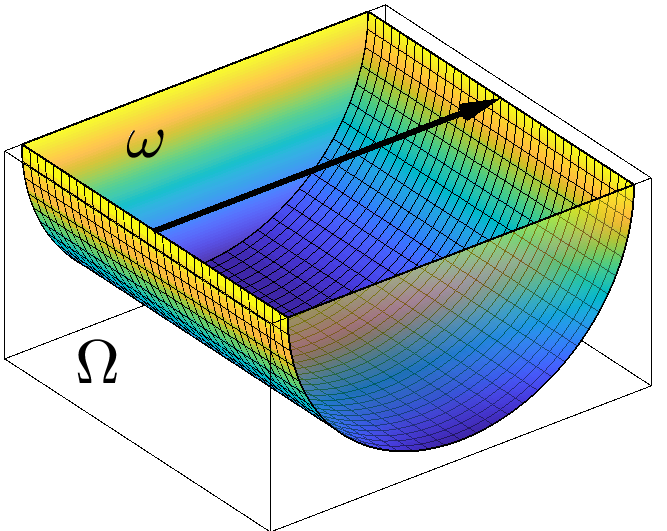}  
\end{center}
\caption{A semicircular cavity $\omega$ of width 1, depth 1, and height 0.5 (embedded into a fictitious domain $\Omega$) where the radius of circular sector is 0.5.}\label{fig.6}
\end{figure}
\begin{figure}[t!]
\begin{center}
\leavevmode
\includegraphics[width=4.5in]{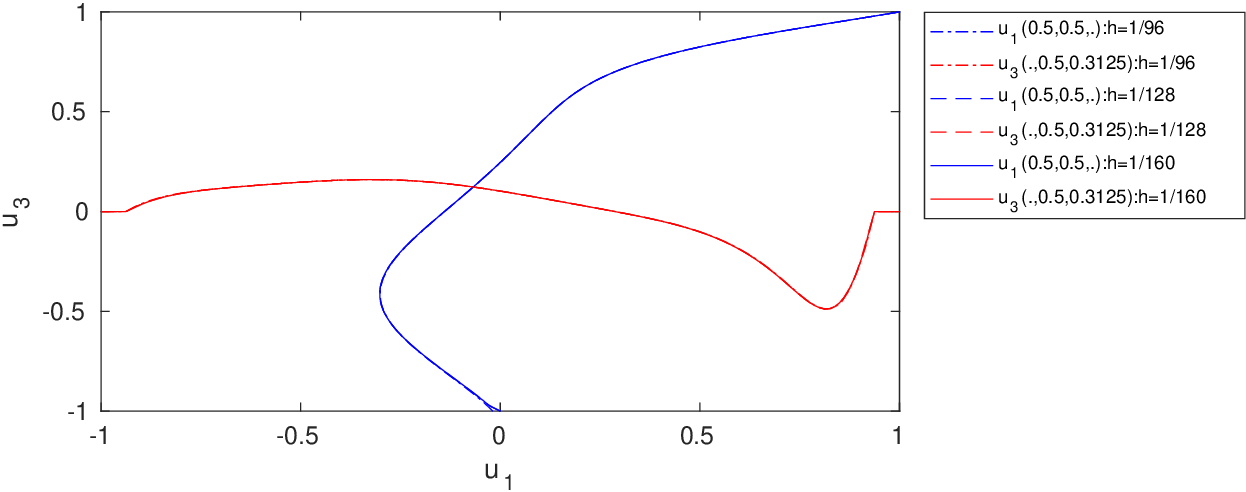}\\
\begin{minipage}{0.49\textwidth}
\includegraphics[width=0.38in]{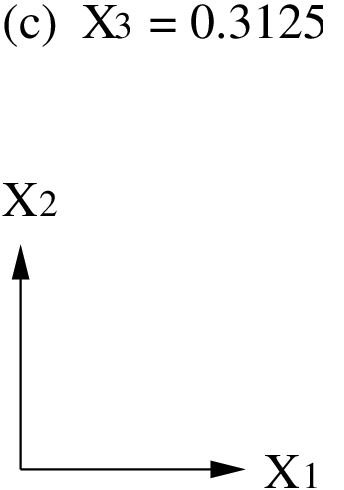}\ 
\includegraphics[width=2.in] {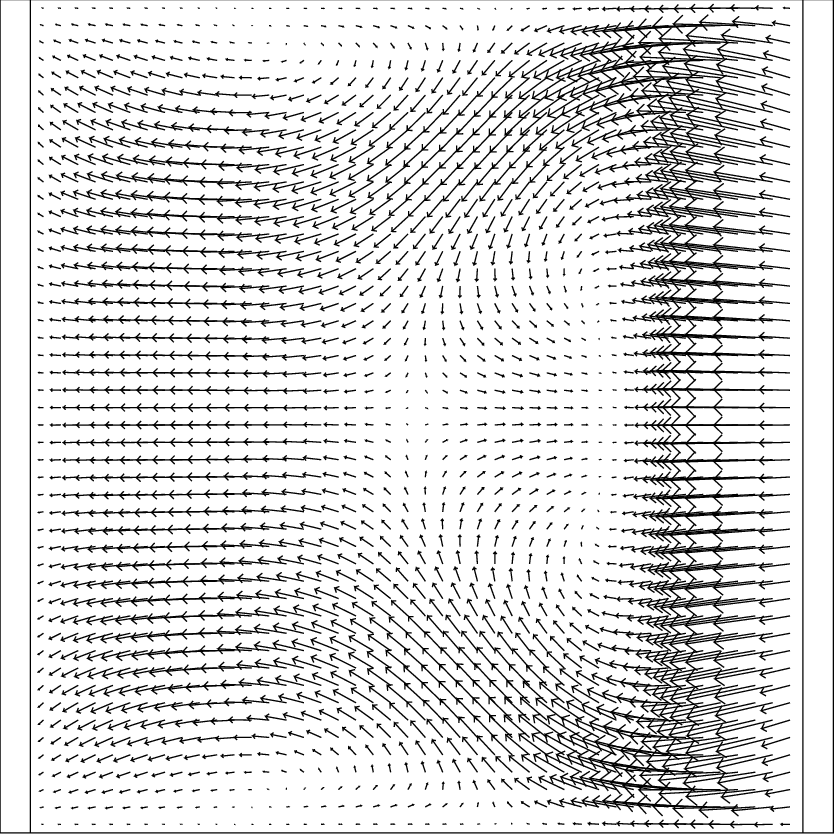}
\end{minipage}
\hskip -5pt
\begin{minipage}{0.49\textwidth}
\includegraphics[width=0.38in]{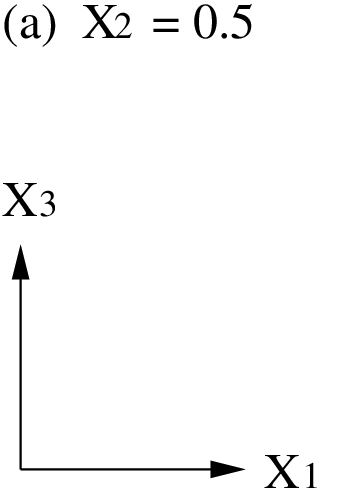}\
\includegraphics[width=2.1in] {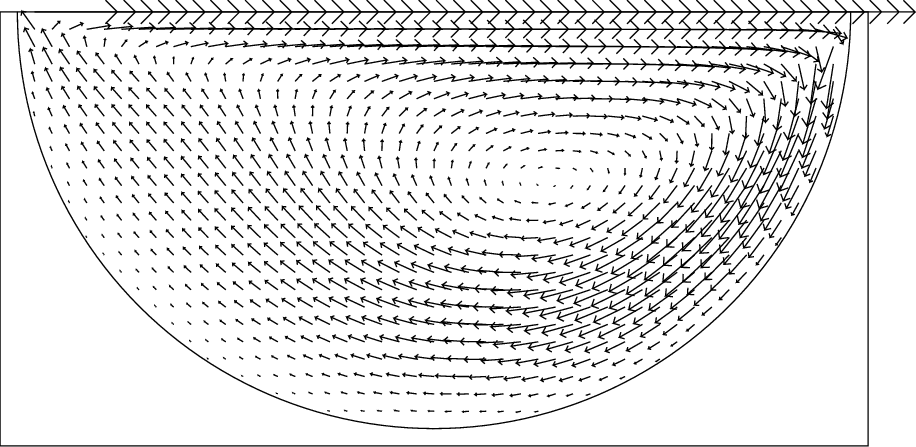}\\
\includegraphics[width=0.38in]{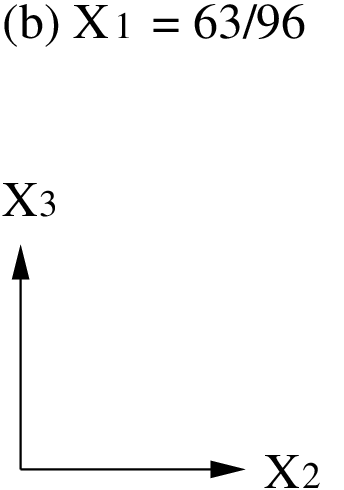}\ 
\includegraphics[width=2.in] {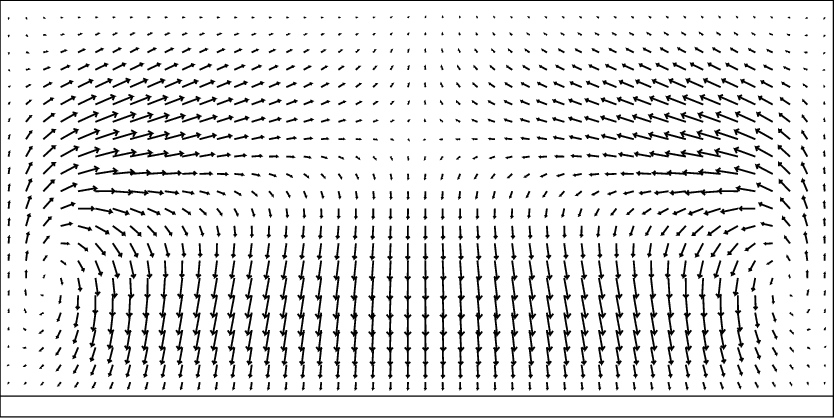}
\end{minipage}	
\end{center}
\caption{(i) Re=400: Comparisons of the  numerical results obtained for $h=1/96$, 1/128, and 1/160 (top),
(ii) Steady flow  velocity vector projected on 
planes: $x_3=0.3125$ (bottom left),  $x_2=0.5$ (middle right), and $x_1=63/96$ (bottom right)
for  $h=1/96$ and   $\triangle t$=0.001. (In the bottom left and right plots,
vector scale is two times of the actual one to enhance visibility.)}\label{fig.7}
\end{figure}

\begin{figure}[t!]
\begin{center}	
\leavevmode
\includegraphics[width=4.5in]{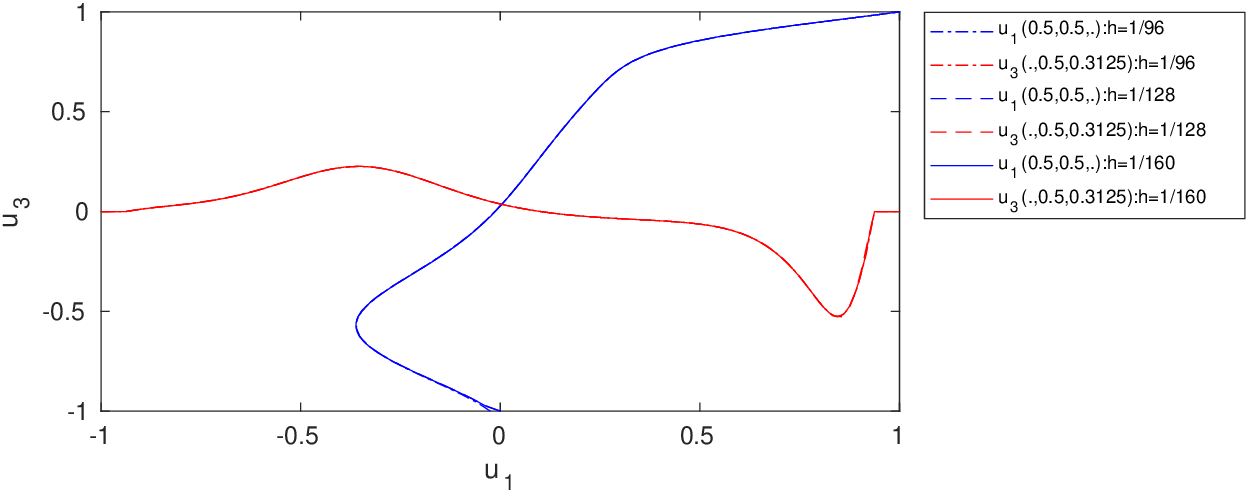}\\
\begin{minipage}{0.49\textwidth}
\includegraphics[width=0.38in]{cord-c1.eps}\ 
\includegraphics[width=2.in] {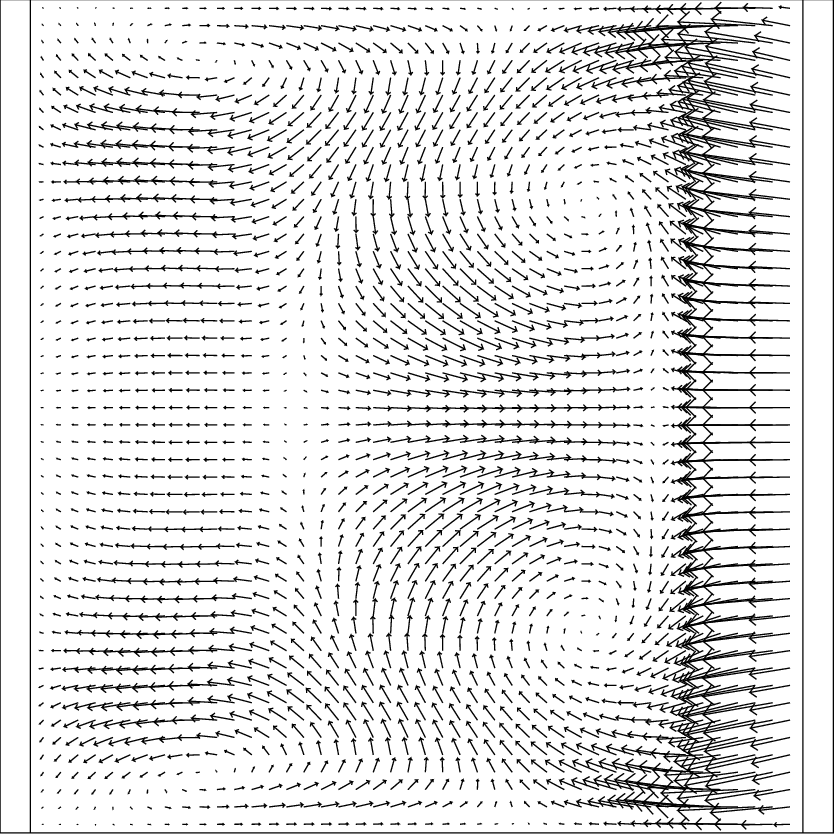}
\end{minipage}
\hskip -5pt
\begin{minipage}{0.49\textwidth}
\includegraphics[width=0.38in]{cord-a1.eps}\
\includegraphics[width=2.1in] {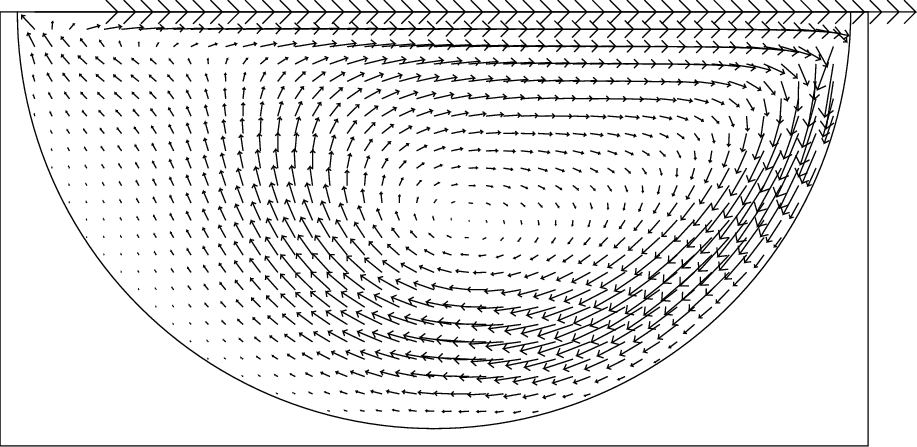}\\
\includegraphics[width=0.38in]{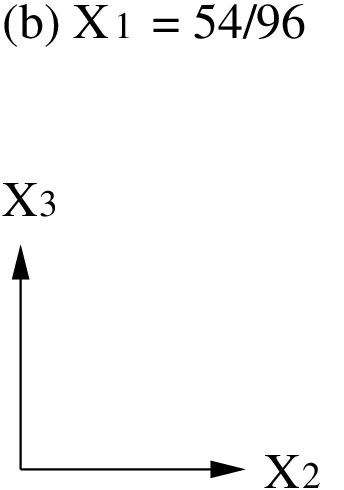}\ 
\includegraphics[width=2.in] {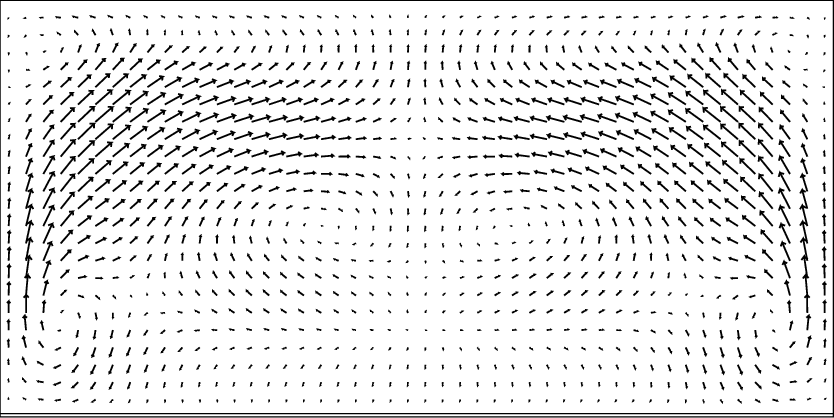}
\end{minipage}	
\end{center}
\caption{(i) Re=900: Comparisons of the  numerical results obtained for $h=1/96$, 1/128, and 1/160 (top),
(ii) Steady flow  velocity vector  projected on 
planes: $x_3=0.3125$ (bottom left),  $x_2=0.5$ (middle right), and $x_1=54/96$ (bottom right)
for  $h=1/96$ and   $\triangle t$=0.001. (In the bottom left and right plots,
vector scale is two times of the actual one to enhance visibility.)}\label{fig.8}
\end{figure}

For  the first semicircular  cavity  shown in Figure \ref{fig.3}, we have taken 
$\Omega = (-2h,1+2h)\times(0,2)\times(-2h,0.5)$  as a computational domain (fictitious domain) 
where $h$ is the mesh size for the velocity field and the radius of circular sector  is 0.5 
and defined the Dirichlet data  $\bU_B$ by
\begin{equation}
	\bU_B(\bx) = \begin{cases} (1,0,0)^T \ \text{on} \ \{\bx \ | \ \bx = (x_1,x_2, 0.5)^T, 0 < x_1< 1, 0<x_2 < 2\},\\
		{\bf 0} \ \text{elsewhere \ on} \ \partial\Omega,
	\end{cases}\label{eqn:3.1}
\end{equation}
Then the Reynolds number is Re=$1/\nu$.  We assumed that a steady state has been reached  when the  change
between two consecutive time steps, $\|\bU^n_h - \bU^{n-1}_h\|_{\infty} /{\triangle t}$,
in the simulation is less than $10^{-7}$,  and then took $\bU_h^n$ as the steady  state solution.   
This semicircular cavity was one of several cavity shapes used in \cite{Migeon2000} to study the shape 
influence on birth and evolution of recirculating flow structures in cavities. The Reynolds number for the 
experiment considered in   \cite{Migeon2000} is Re=1000. To validate the numerical methodologies briefly 
described in the previous section, we have obtained the initial phase of the flow establishment with the 
velocity mesh size $h=1/96$  and time step $\triangle t$=0.001.  For the star-up flow, 
numerical results were computed for the first 9 seconds (see Figure \ref{fig.4}). In Figure. \ref{fig.5}, 
the vortex-core trajectory of numerical results on the middle vertical plane $x_2=1$ shows a very good agreement  
with  those obtained in  \cite{Migeon2000}. Our numerical results suggest that the critical Reynolds number for
the transition from steady state flow to oscillatory one is close to Re=790. The critical Reynolds number
suggested in  \cite{Migeon2000} is higher than 1000 since they believed that  the  lid-driven flow in 
a semicircular cavity reported in  \cite{Migeon2000}  is steady for Re=1000.

To investigate the effect of semicircular shape on the transition of  lid-driven cavity flows, 
we have considered the second semicircular  cavity  as shown in Figure \ref{fig.6}.
Its radius of circular sector  is 0.5 and fictitious domain is
$\Omega = (-2h,1+2h)\times(0,1)\times(-2h,0.5)$.
The Dirichlet data  $\bU_B$ is defined as
\begin{equation}
	\bU_B(\bx) = \begin{cases} (1,0,0)^T \ \text{on} \ \{\bx \ | \ \bx = (x_1,x_2, 0.5)^T, 0 < x_1, x_2 < 1\},\\
		{\bf 0} \ \text{elsewhere \ on} \ \partial\Omega,
	\end{cases}\label{eqn:3.2}
\end{equation}
We like to compare the resulting flows in this semicircular cavity with those in a shallow cavity 
with a unit square base and height 0.5 discussed in \cite{PanChiuGuoHe2023}.

\begin{figure}[t!]
\begin{center}
\leavevmode 
\hskip 10pt \includegraphics[width=3.5in]{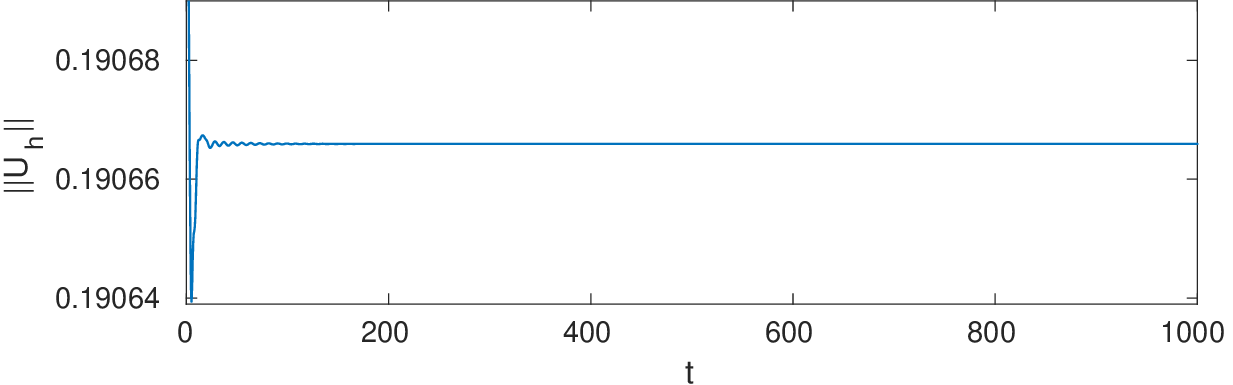}\\  
	 		 \includegraphics[width=3.8in]{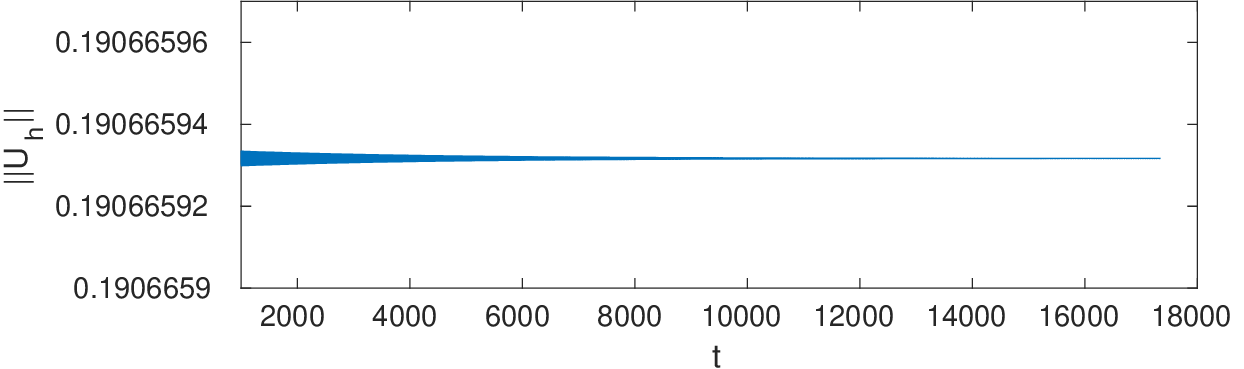} \\ 	 		 
\hskip 10pt \includegraphics[width=3.5in]{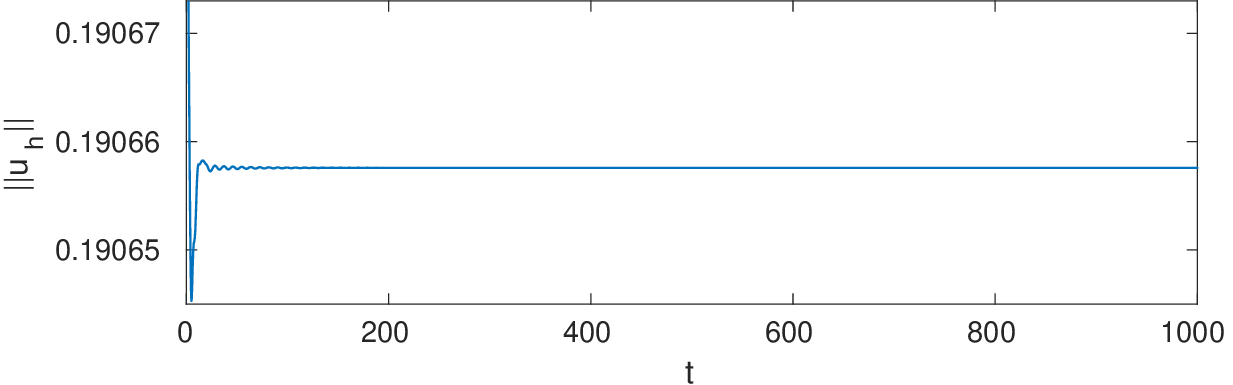} \\ 
	 		 \includegraphics[width=3.8in]{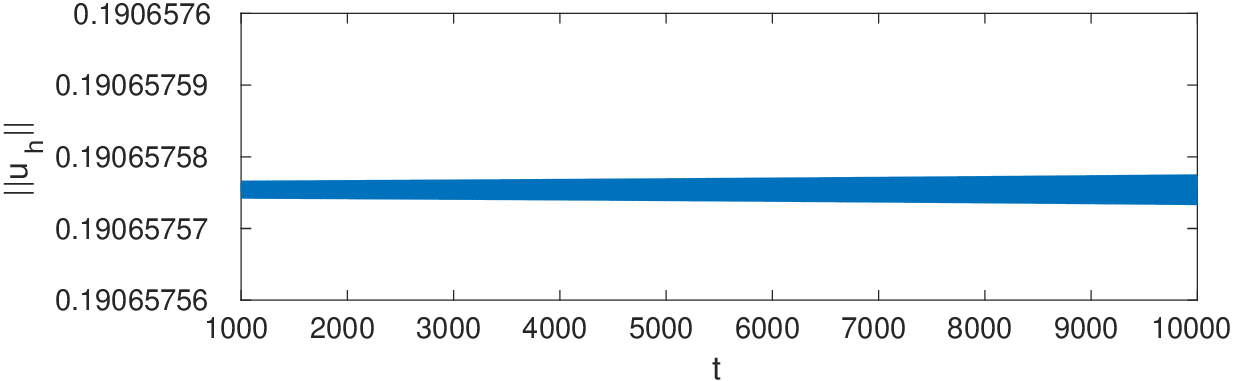}  	 		 
\end{center}
\caption{Histories of $\|\bu_h\|$ for $Re=927$  (top two) and 928 (bottom two) obtained with  
$h=1/96$ and $\triangle t$=0.001. } \label{fig.9}
\end{figure}

\begin{figure} [!t]
\begin{center}
\leavevmode
\includegraphics[width=4.75in]{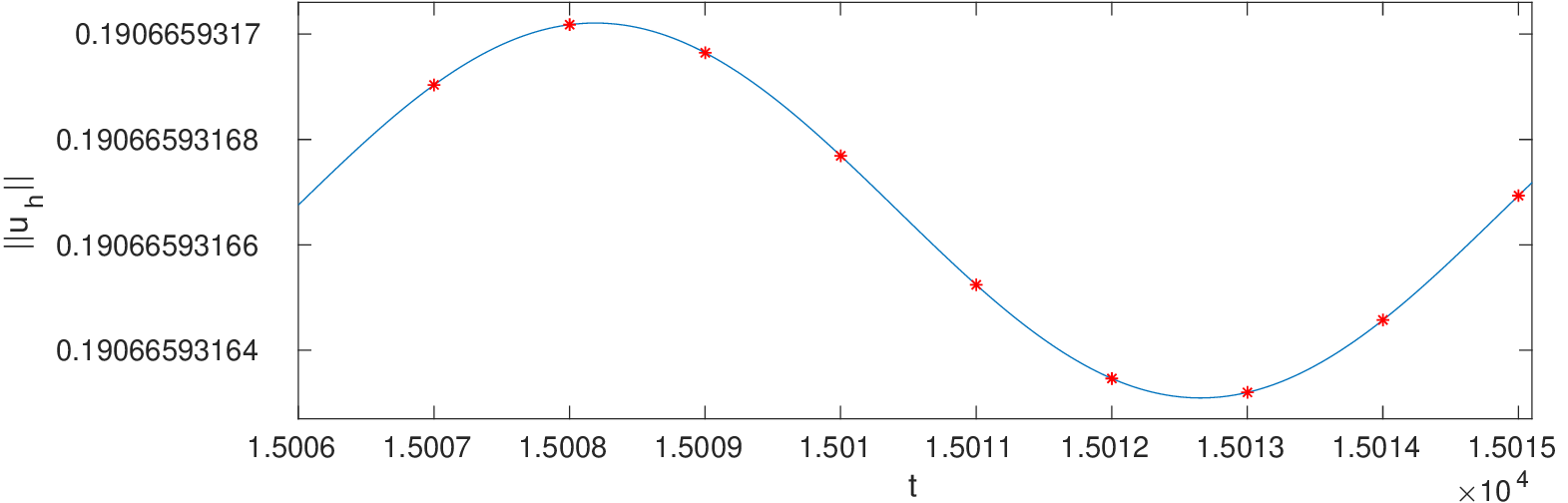}\\
\begin{minipage}{0.49\textwidth}
	\includegraphics[width=0.38in]{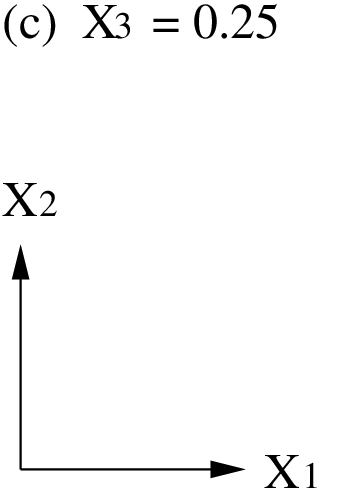}\ 
	\includegraphics[width=2.1in] {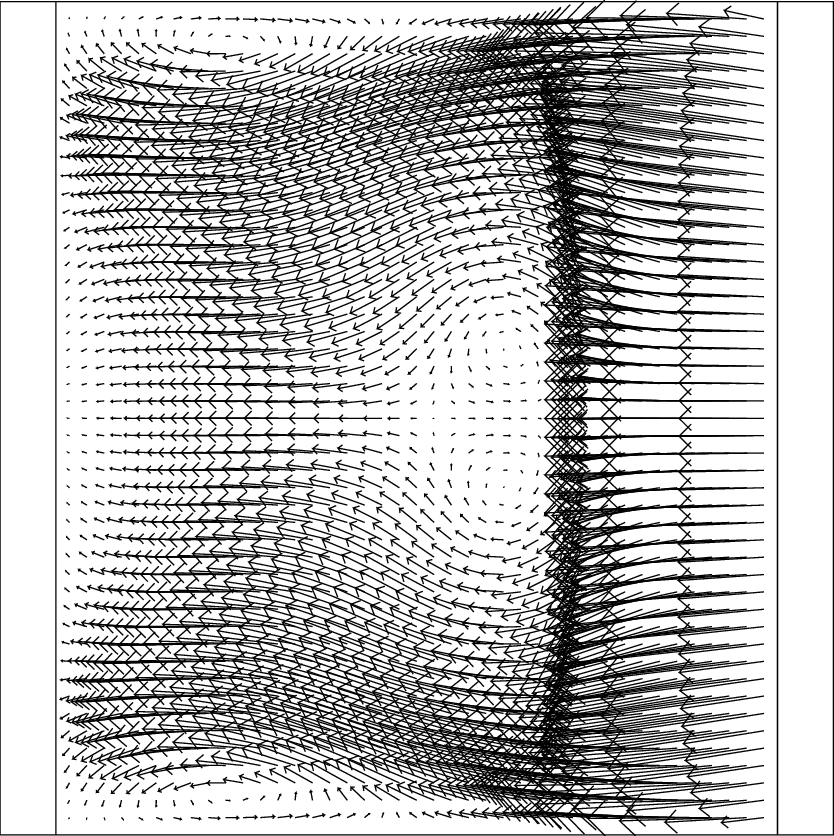}
\end{minipage}
\hskip -5pt
\begin{minipage}{0.49\textwidth}
	\includegraphics[width=0.38in]{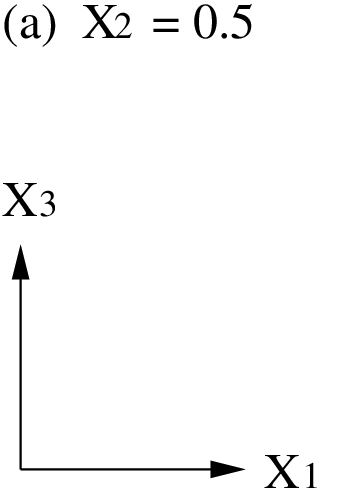}\
	\includegraphics[width=2.2in] {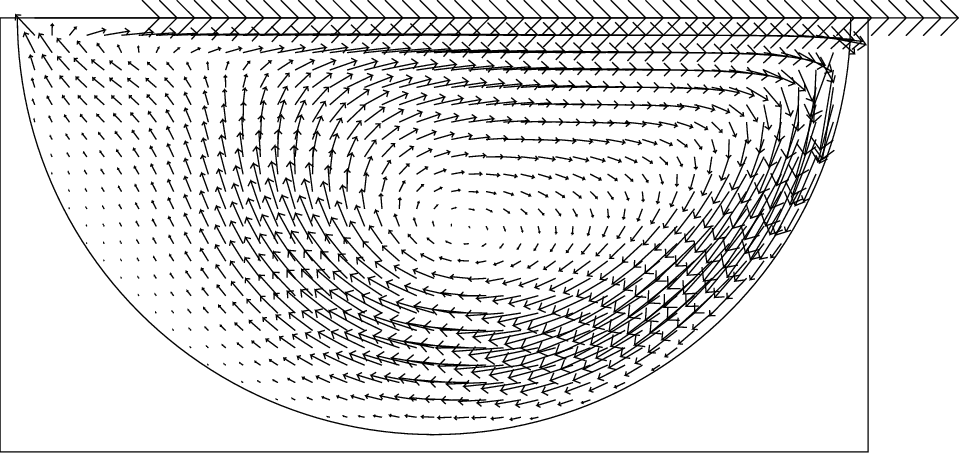}\\
	\includegraphics[width=0.38in]{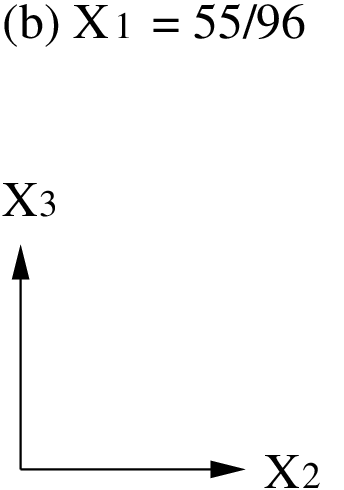}\ 
	\includegraphics[width=2.1in] {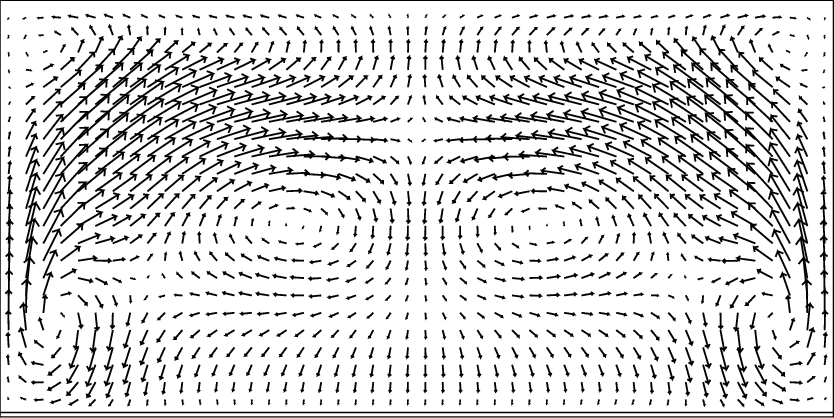} 
\end{minipage}
\end{center}
\caption{History of $\|\bu_h\|$ for one period in a semicircular cavity for  $Re=927$ (top), 
averaged velocity field projected one planes: $x_3=0.25$ (bottom left),  $x_2=0.5$ (middle right), 
and $x_1=55/96$ (bottom right) for  $h=1/96$ and   $\triangle t$=0.001. (In the bottom left and right 
plots, the vector scale is four times that of the actual one to enhance visibility.)}\label{fig.10}
\end{figure}
\begin{figure} 
\begin{center}
\leavevmode
\hskip -8pt \includegraphics[width=0.38in]{cord-b3.eps}\
\includegraphics[width=2.0in]{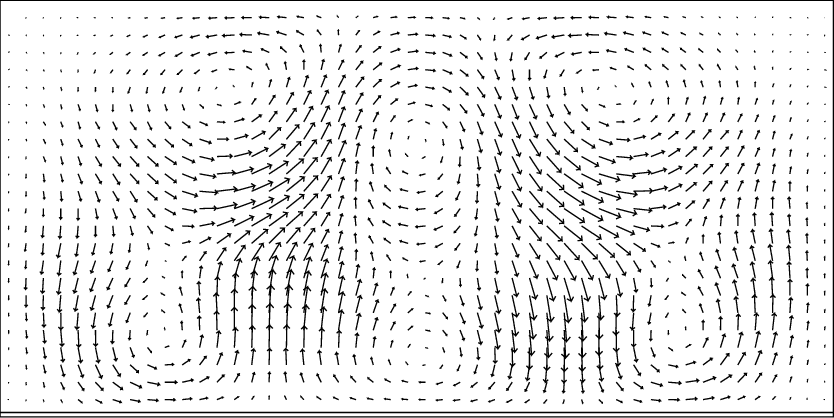} \  \includegraphics[width=2.0in]{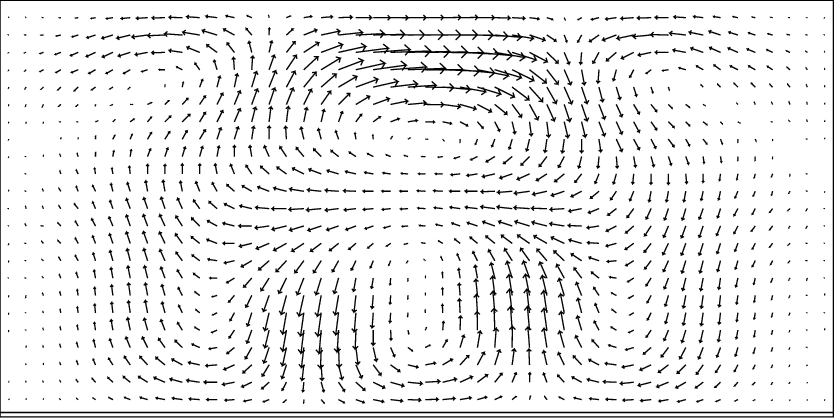} \\
\hskip 22pt \includegraphics[width=2.0in]{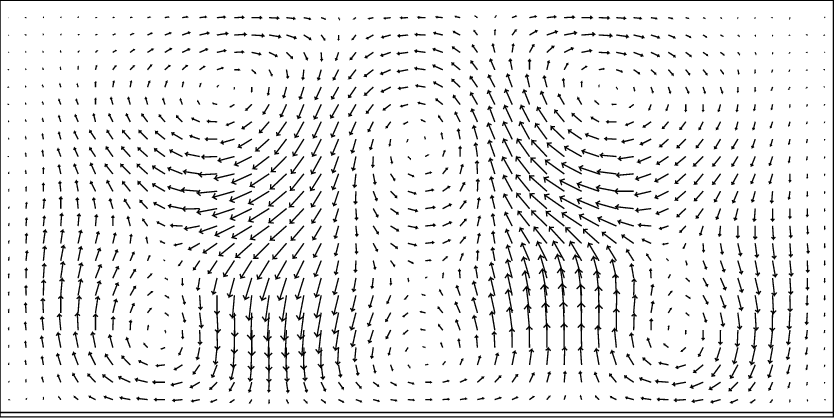} \  \includegraphics[width=2.0in]{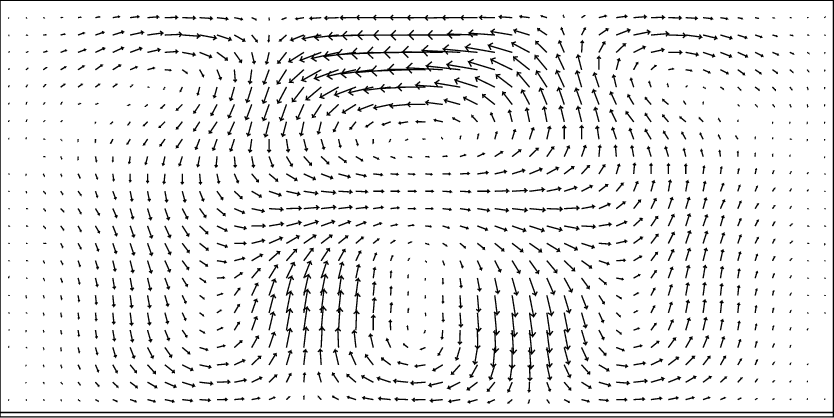} \\
\hskip -8pt \includegraphics[width=0.38in]{cord-c3.eps}\
\includegraphics[width=2.0in]{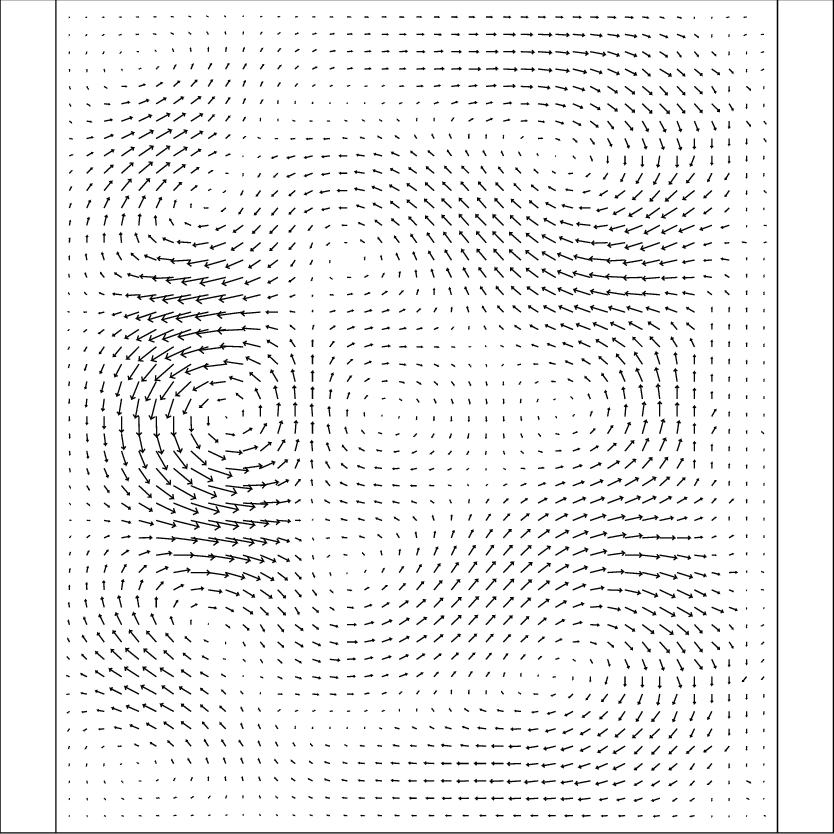} \  \includegraphics[width=2.0in]{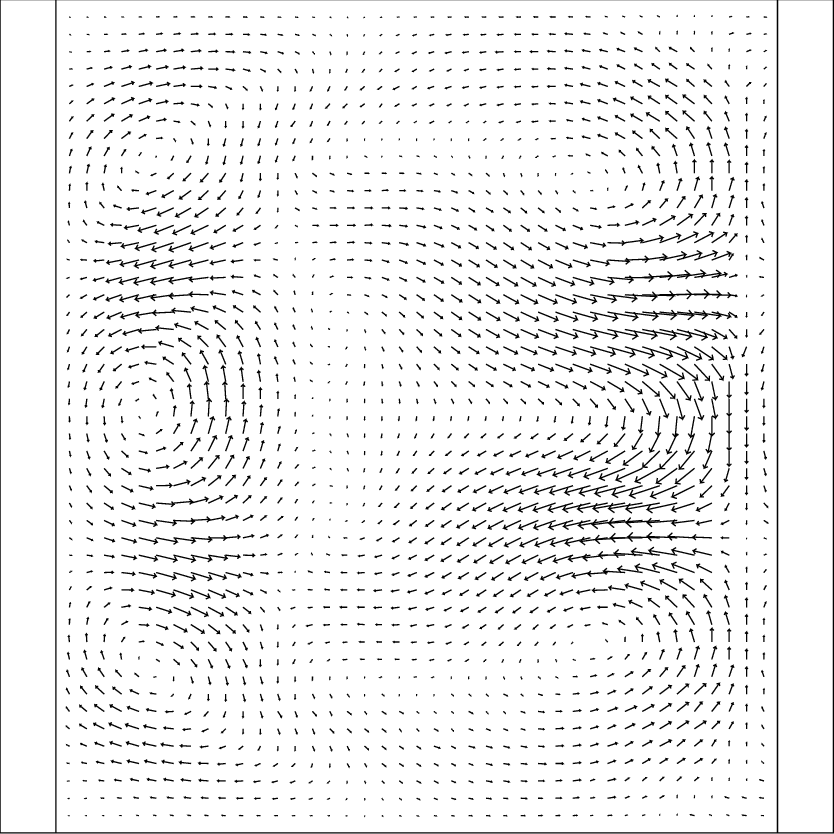} \\
\hskip 24pt \includegraphics[width=2.0in]{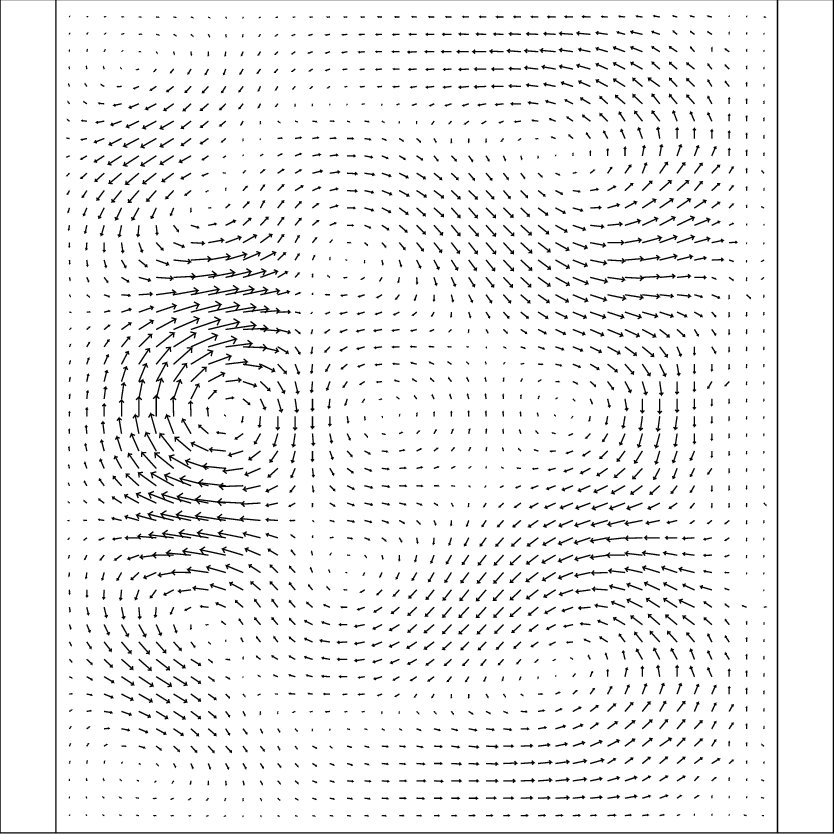} \  \includegraphics[width=2.0in]{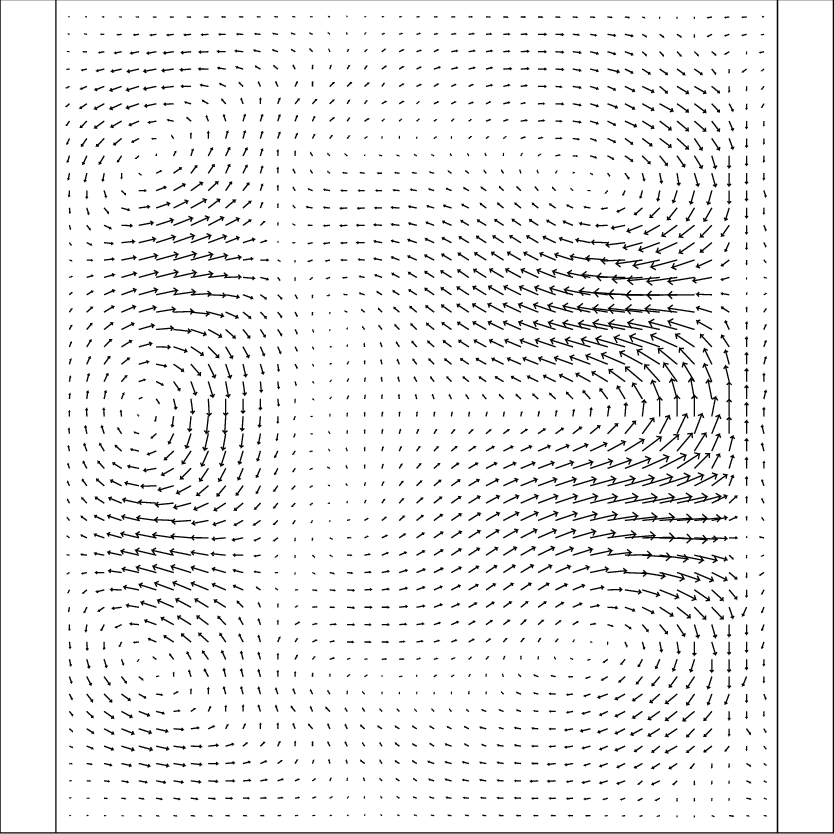} 
\end{center}
\caption{Velocity field oscillation (difference) projecting on the planes $x_1=55/96$ (top four) and $x_3=0.25$ (bottom four)
at $t=15008$, 15010.5, 15012.5, and 15015 (from left to right and then from top to bottom) for Re=927. 
The velocity vectors have been magnified  in those  plots to enhance visibility.}\label{fig.11}
\end{figure}

The steady flow velocity vectors for Re=400 and 900 are shown in Figures \ref{fig.7} and \ref{fig.8},
respectively where the velocity field vectors for Re=400 (resp., Re=900) are projected onto the three planes, 
$x_2 = 0.5$,  $x_1 = 63/96$, and $x_3 = 0.3125$ (resp., $x_2 = 0.5$,  $x_1 =54/96$, and $x_3 = 0.3125$). 
The length of vectors has been enlarged two times in the two planes, $x_1 =54/96$ and $x_3 = 0.3125$, to improve clarity. 
The two plots on $x_2=0.5$ show that the center of primary vortex moves toward the central region as Re increases 
from 400 to 900.  At Re=400, on $x_1 = 63/96$,  there is a vortex occurring close to each bottom corner but
another  pair of vortices is shown in the central region on $x_1=54/96$ for Re=900. Similarly, 
four established vortices appear on  $x_3 = 0.3125$ for Re=900; but not for Re=400.

For lid-driven flows in a cavity, one of many interesting questions is to locate the critical Reynolds 
number Re$_{cr}$ for a transition from a steady lid-driven flow to oscillatory one. In, e.g., \cite{Feldman2010} 
and \cite{Liberzon2011},  Re$_{cr}$ was predicted to be less than 2000  for lid-driven flows in a cubic cavity.  
On the other hand, Gianetti {\it et al.}  found (ref. \cite{Giannetti2009}) that the cubic  lid-driven 
cavity flow becomes unstable for Re just  above 2000 via  a global linear stability analysis.  
Kuhlmann and Albensoeder \cite{Kuhlmann2014} obtained numerically 
that the critical Reynolds number value is 1919.51. In \cite{PanChiuGuoHe2023}, Pan {\it et al.}  found
the critical Reynolds number value is between 1894 and 1895 in a cubic cavity. 
These results indicate  that the Hopf bifurcation related  to the  oscillating flows occurs for Re slightly 
below 2000 for lid-driven flows in a cubic cavity.  To study the effect of cavity shape on the transition of 
lid-driven flows, we first like to locate the  value of  Re$_{cr}$ for lid-driven 
flows in a semicircular cavity (see Figure \ref{fig.6}) and then study the oscillation flows 
for the Reynolds numbers close to Re$_{cr}$.
We have computed the flow velocity $\bu^n_h$ for different Re values and mesh sizes $h$ and 
analyzed its history of  $L^2$-norm (i.e., plot of $\|\bu^n_h\|$ versus $t$).
For $h=1/96$ and  $\triangle t$=0.001,  the flow field evolves to a steady state  and the amplitude of 
its $L^2$-norm  oscillation  decreases  in time for Re $ \le 927$ (see  Figure  \ref{fig.9}). 
For Re $\ge 928$,  the steady state criterion is not satisfied and the amplitude of oscillation 
increases in time (see  Figure  \ref{fig.9}). Thus we conclude that the critical 
Reynolds number Re$_{cr}$ for the occurrence of transition is somewhere between 927 and 928.  
The oscillating angular frequency is between 0.70376 and 0.70345. 
Applying  the same analysis to the histories of flow velocity  $L^2$-norm for 
$h=1/160$ and  $\triangle t$=0.001,  the critical Re$_{cr}$ is between 934 and 935. 
The oscillating angular frequency is between 0.70645 and 0.70629.  
\begin{figure} [!t]
	\begin{center}
		\leavevmode
		\includegraphics[width=4.75in]{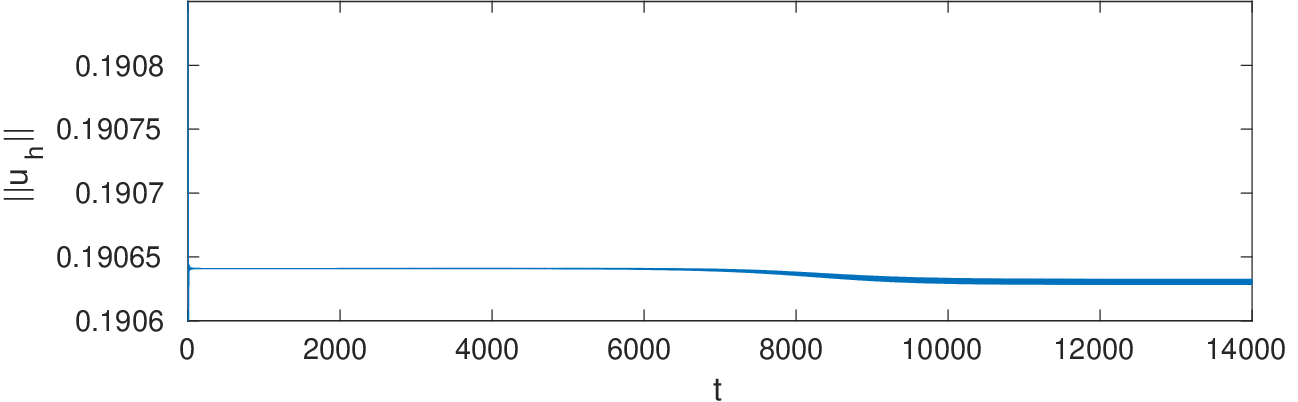}\\
		\includegraphics[width=4.75in]{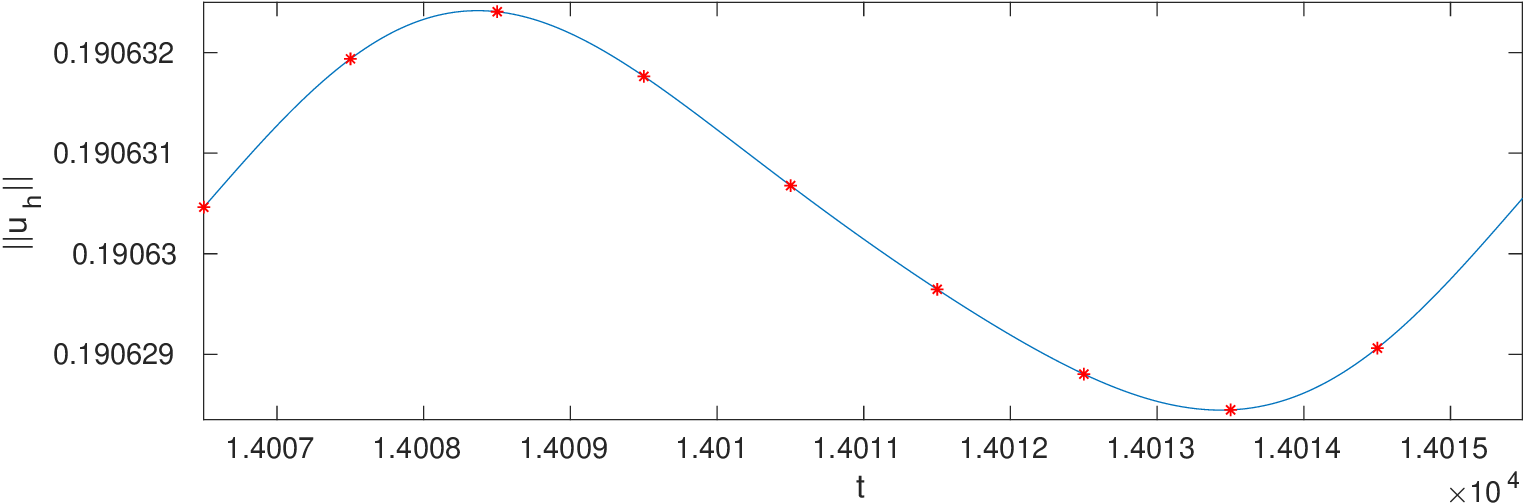}\\
		\begin{minipage}{0.49\textwidth}
			\includegraphics[width=0.38in]{cord-c3.eps}\ 
			\includegraphics[width=2.1in] {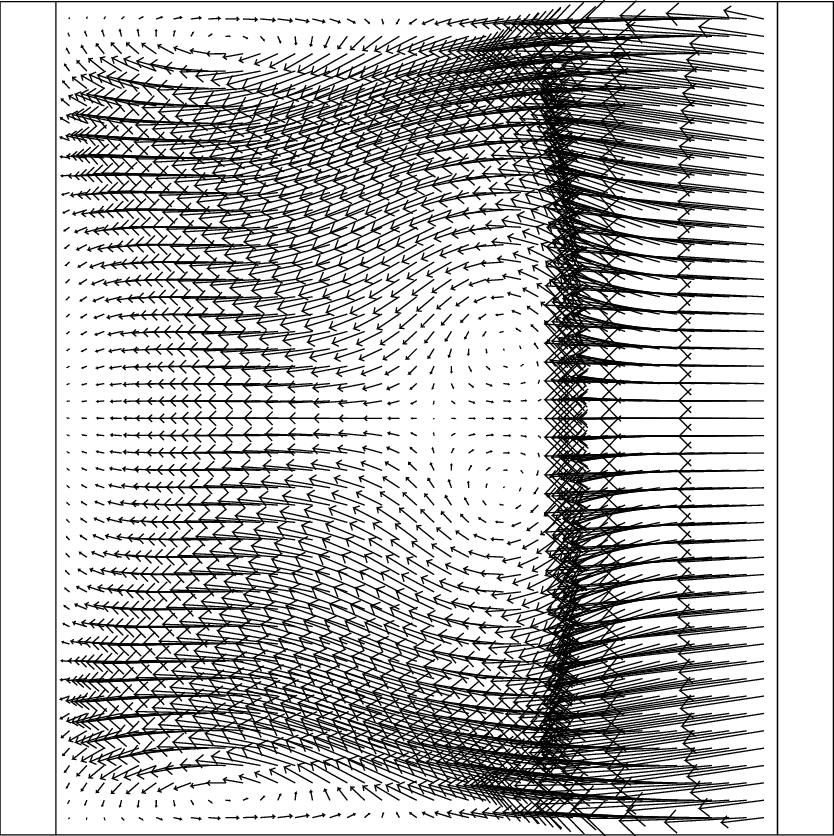}
		\end{minipage}
		\hskip -5pt
		\begin{minipage}{0.49\textwidth}
			\includegraphics[width=0.38in]{cord-a3.eps}\
			\includegraphics[width=2.2in] {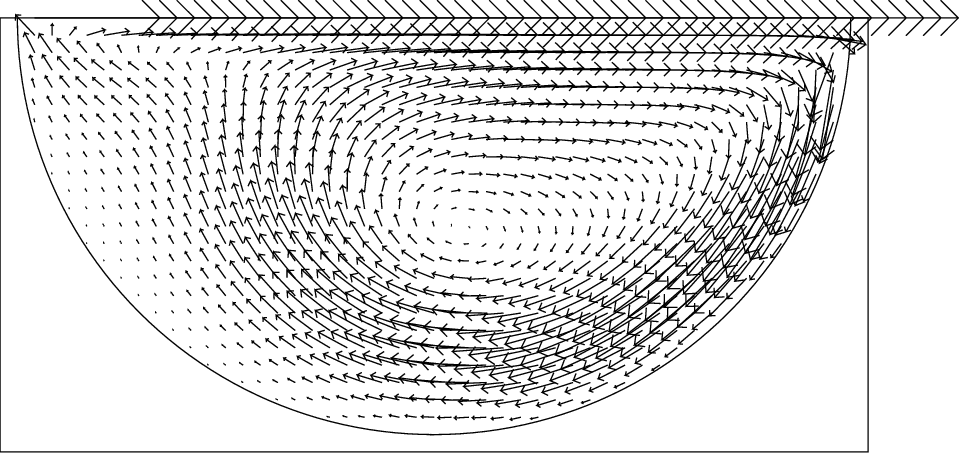}\\
			\includegraphics[width=0.38in]{cord-b3.eps}\ 
			\includegraphics[width=2.1in] {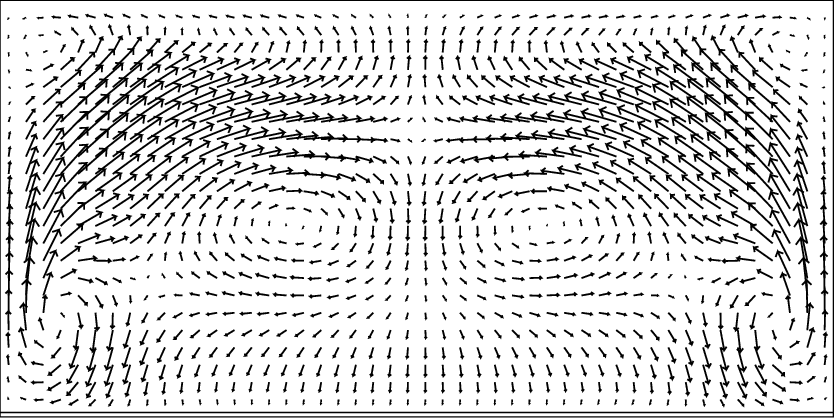} 
		\end{minipage}
	\end{center}
	\caption{History of $\|\bu_h\|$ (top) and that of one period (2nd one from top) in a semicircular cavity 
		for $Re=930$, averaged velocity field projected on  planes: $x_3=0.25$ (bottom left),  $x_2=0.5$ (middle right), 
		and $x_1=55/96$ (bottom right) for  $h=1/96$ and  $\triangle t$=0.001. (In the bottom left and right 
		plots, the vector scale is four times of the actual one to enhance visibility.)}\label{fig.12a}
\end{figure}

\begin{figure} [!t]
\begin{center}
\leavevmode
\includegraphics[width=4.75in]{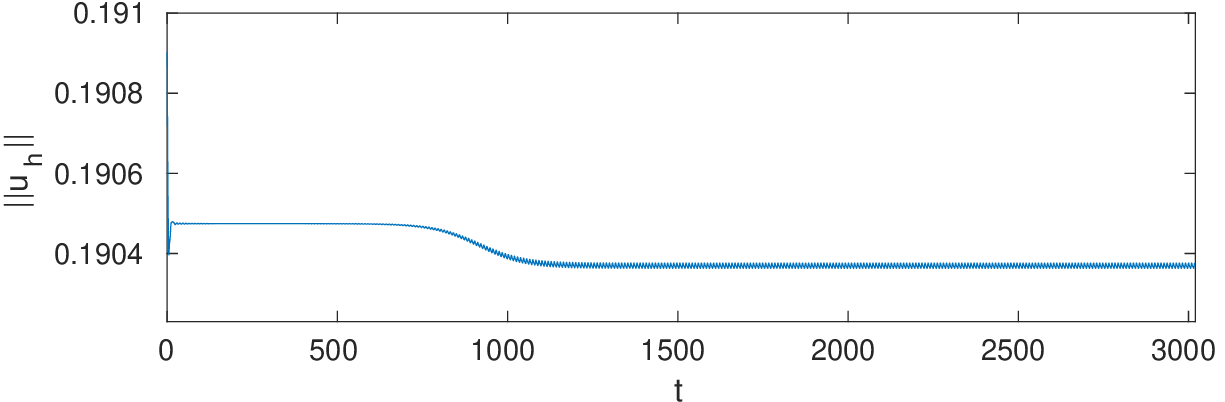}\\
\includegraphics[width=4.75in]{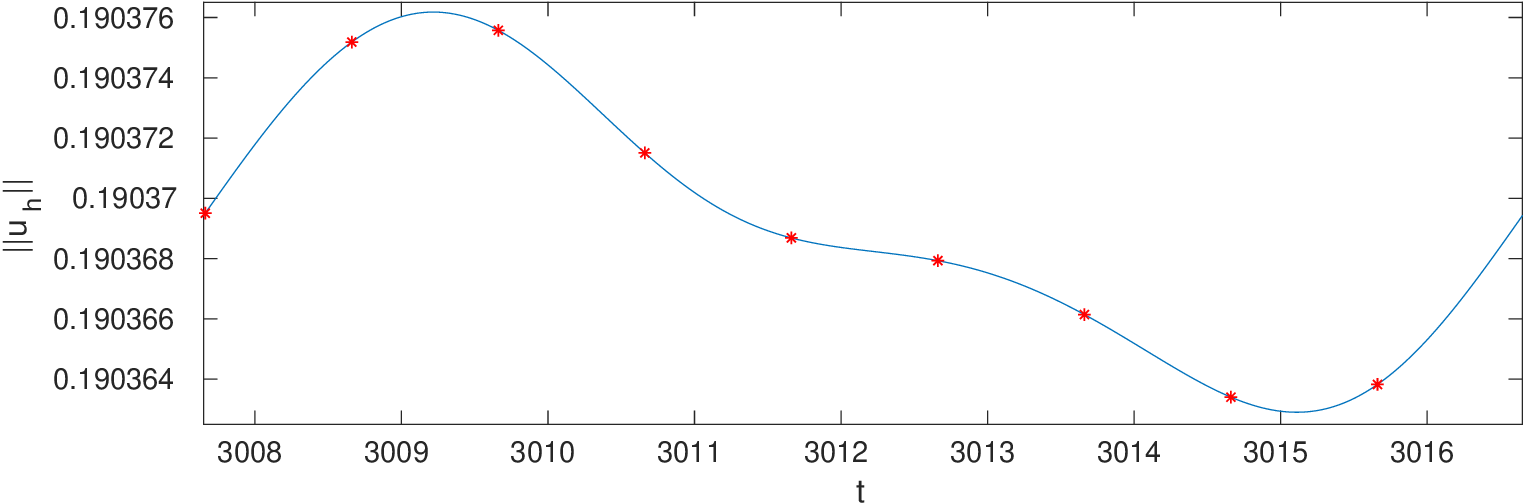}\\
\begin{minipage}{0.49\textwidth}
\includegraphics[width=0.38in]{cord-c3.eps}\ 
\includegraphics[width=2.1in] {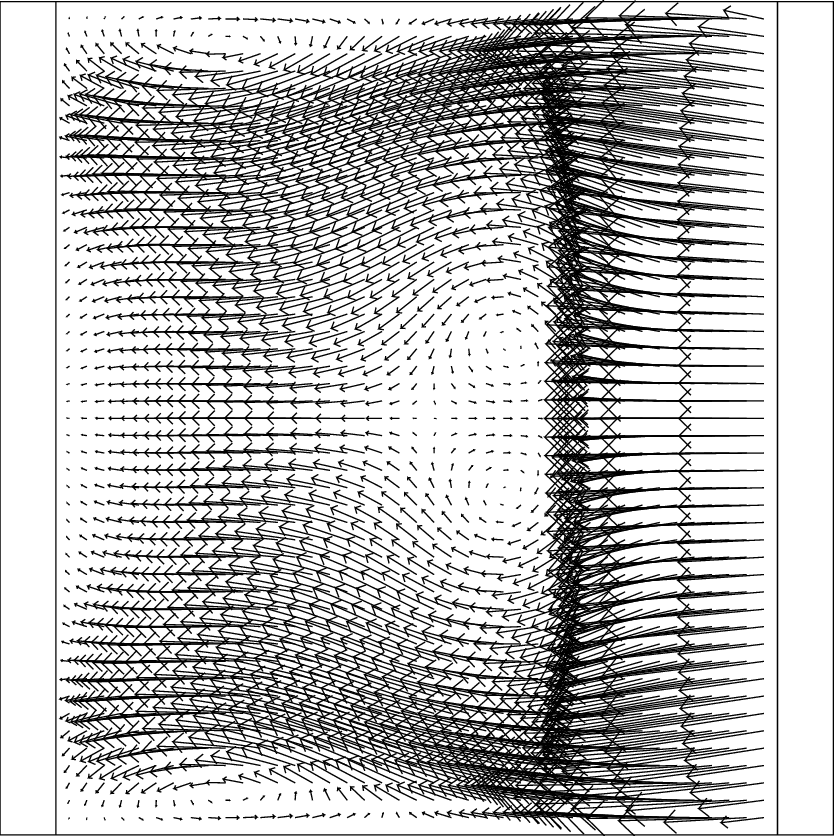}
\end{minipage}
\hskip -5pt
\begin{minipage}{0.49\textwidth}
\includegraphics[width=0.38in]{cord-a3.eps}\
\includegraphics[width=2.2in] {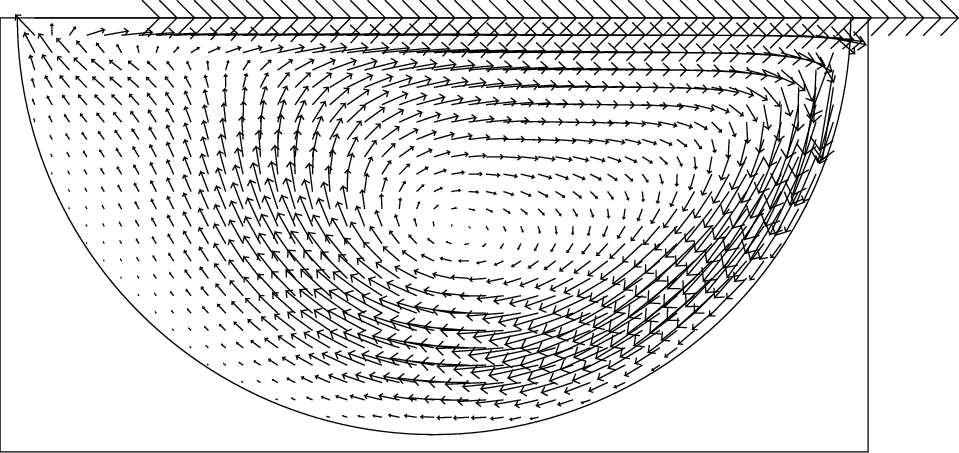}\\
\includegraphics[width=0.38in]{cord-b3.eps}\ 
\includegraphics[width=2.1in] {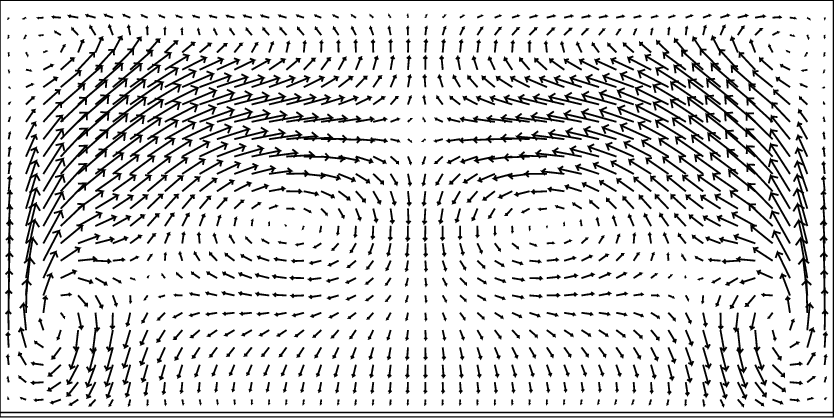} 
\end{minipage}
\end{center}
\caption{History of $\|\bu_h\|$ (top) and that of one period (2nd one from top) in a semicircular cavity 
for $Re=950$, averaged velocity field projected on  planes: $x_3=0.25$ (bottom left),  $x_2=0.5$ (middle right), 
and $x_1=55/96$ (bottom right) for  $h=1/96$ and  $\triangle t$=0.001. (In the bottom left and right 
plots, the vector scale is four times of the actual one to enhance visibility.)}\label{fig.12}
\end{figure}

\begin{figure} [!t]
\begin{center}
\leavevmode
\includegraphics[width=1.73in]{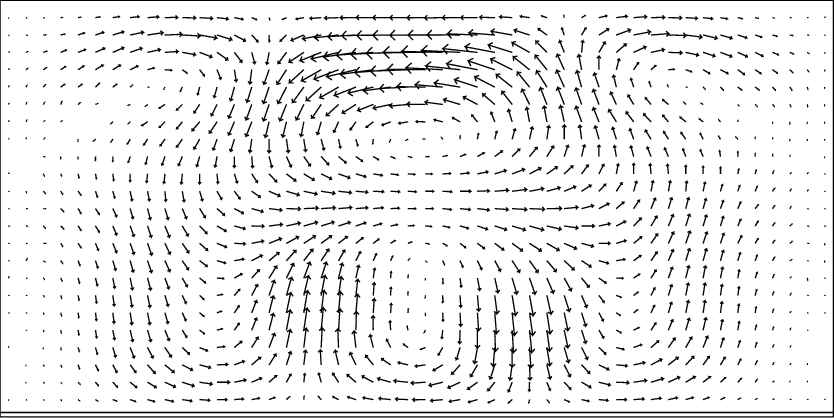}\hskip 2pt
\includegraphics[width=1.73in]{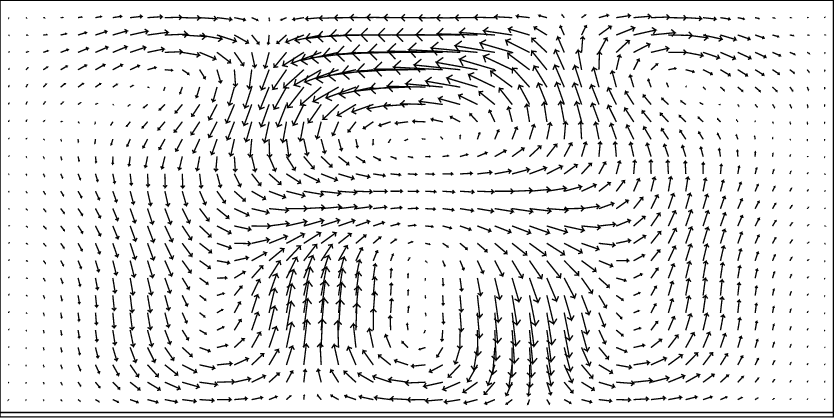}\hskip 2pt
\includegraphics[width=1.73in]{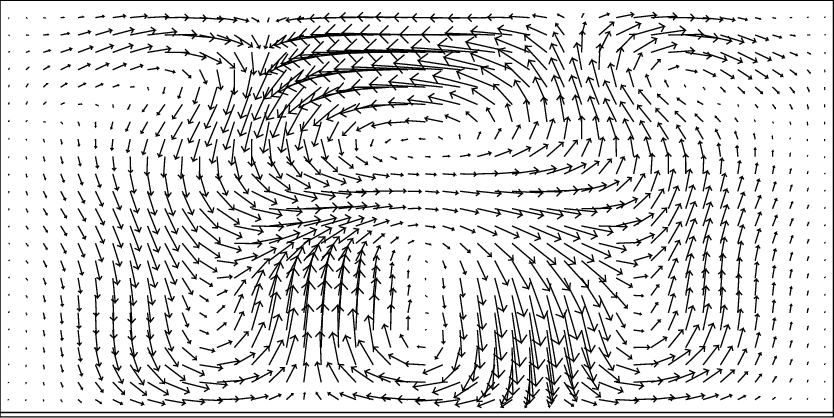} \\
\includegraphics[width=1.73in]{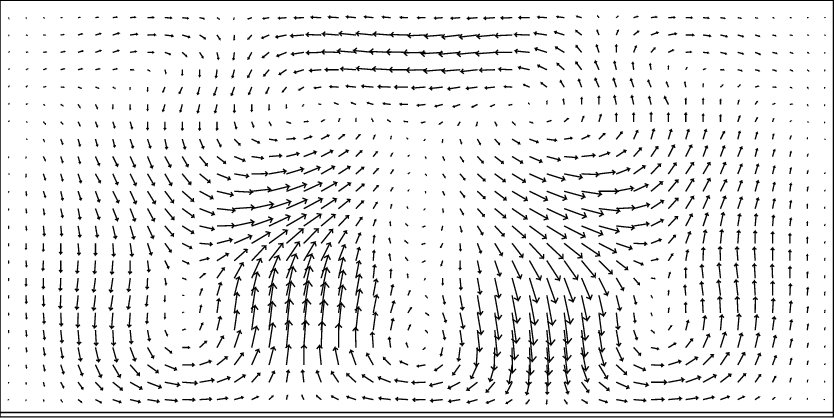}\hskip 2pt
\includegraphics[width=1.73in]{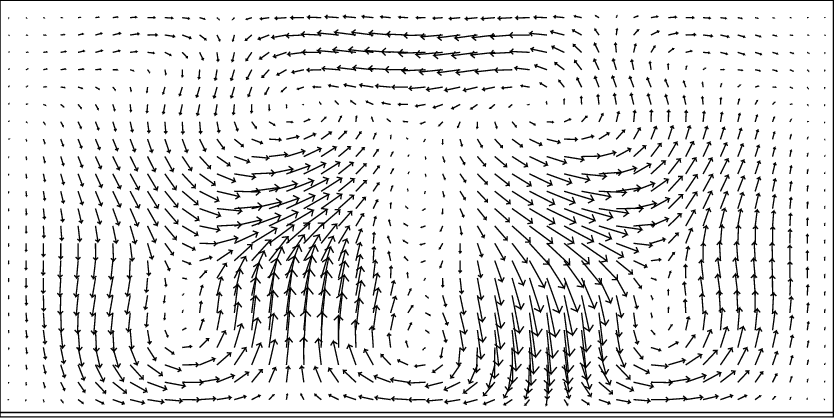}\hskip 2pt \includegraphics[width=1.73in]{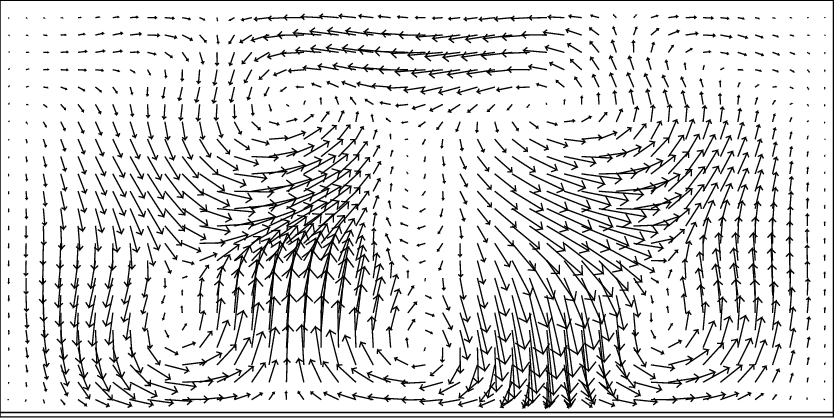} \\
\includegraphics[width=1.73in]{vec-b-circular-square-Re0927-h096-average-t15008-2000000x-iy57.eps}\hskip 2pt
\includegraphics[width=1.73in]{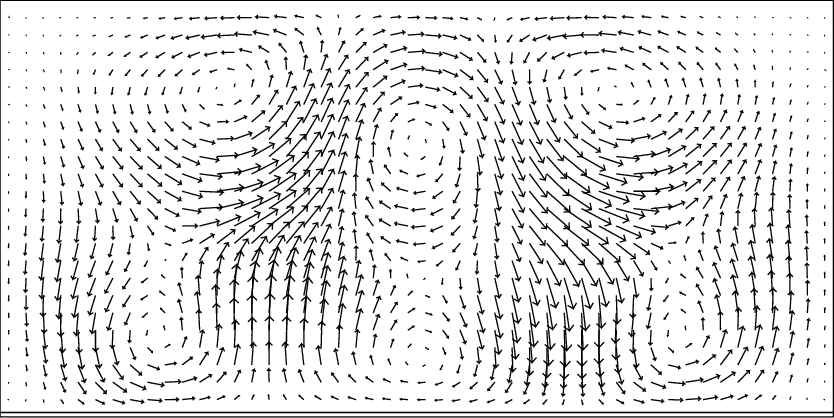}\hskip 2pt
\includegraphics[width=1.73in]{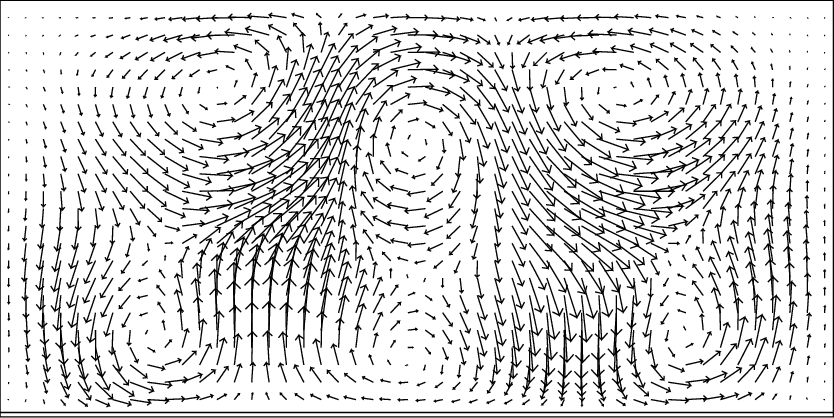} \\
\includegraphics[width=1.73in]{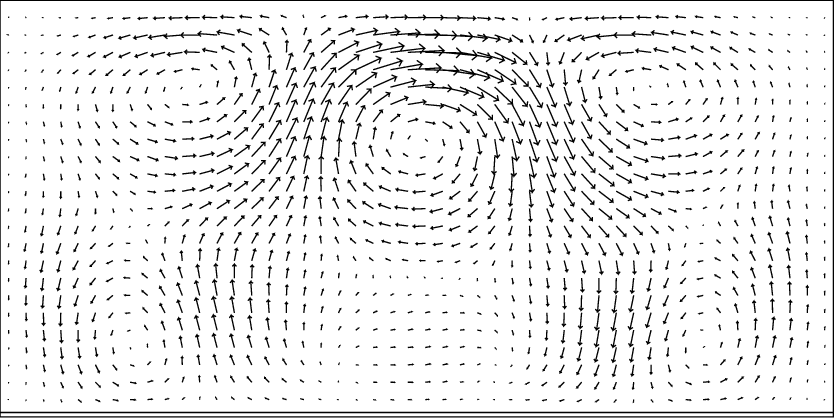}\hskip 2pt
\includegraphics[width=1.73in]{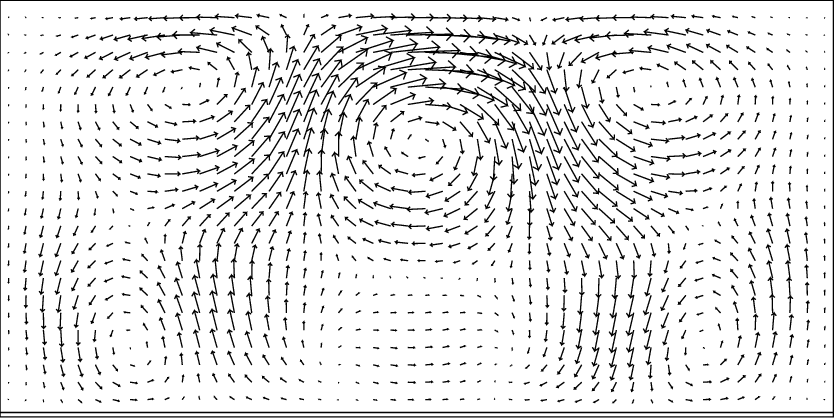}\hskip 2pt
\includegraphics[width=1.73in]{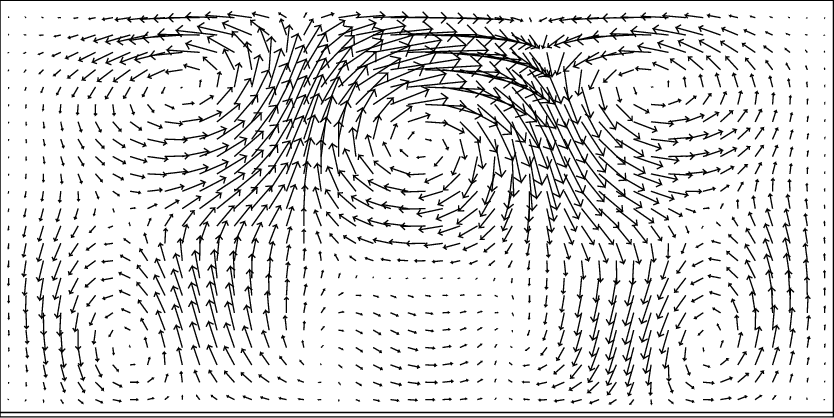} \\
\includegraphics[width=1.73in]{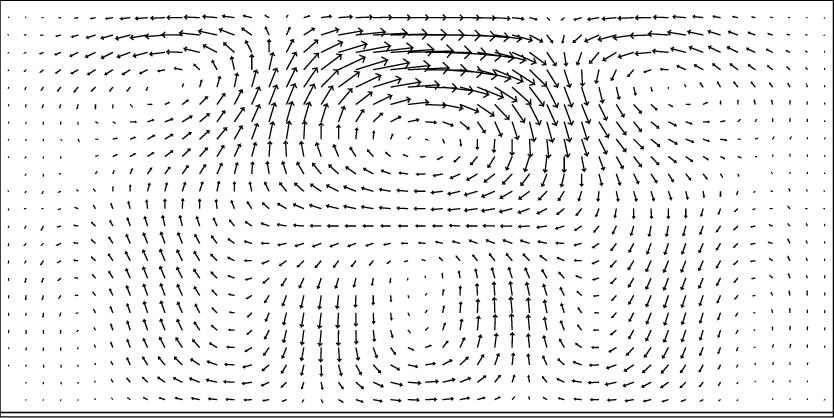}\hskip 2pt
\includegraphics[width=1.73in]{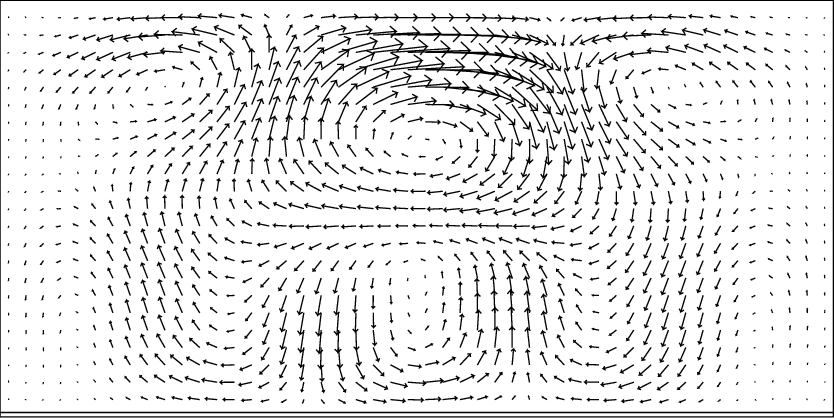}\hskip 2pt
\includegraphics[width=1.73in]{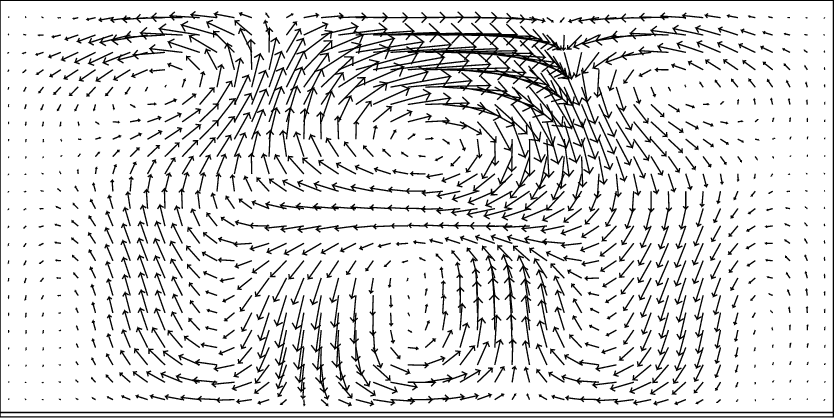} \\
\includegraphics[width=1.73in]{vec-b-circular-square-Re0927-h096-average-t15010d5-2000000x-iy57.eps}\hskip 2pt
\includegraphics[width=1.73in]{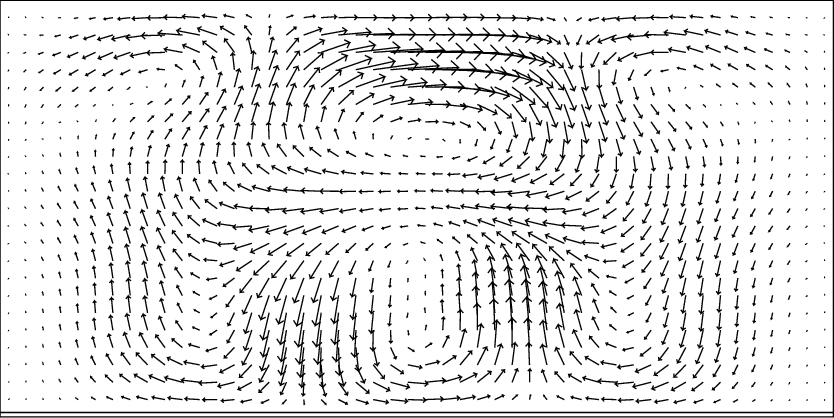}\hskip 2pt
\includegraphics[width=1.73in]{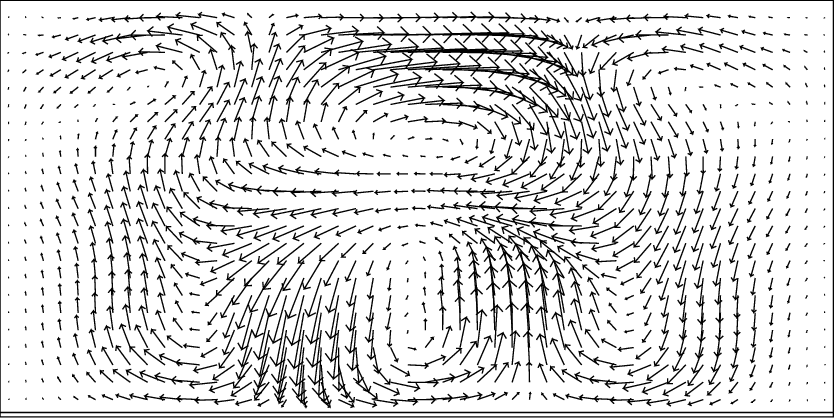} 
\end{center}
\caption{Oscillation mode projecting on  plane $x_1=55/96$ in a semicircular cavity  for Re=927 (left) at 
$t=$15006, 15007, 15008, 15009, 15010, and 15010.5 (from top to bottom),    
Re=930 (middle) at $t=$ 14006.5, 14007.5, 14008.5, 14009.5, 14010.5, and 14011 (from top to bottom),
and Re=950 (right) at $t=$ 3007.65,  3008.65, 3009.65, 3010.65, 3011.65, and 3012.15 (from top to bottom).  
Velocity vectors have been magnified to enhance visibility.}\label{fig.13}
\end{figure} 

\begin{figure} [!t]
\begin{center}
\leavevmode                   
\includegraphics[width=1.73in]{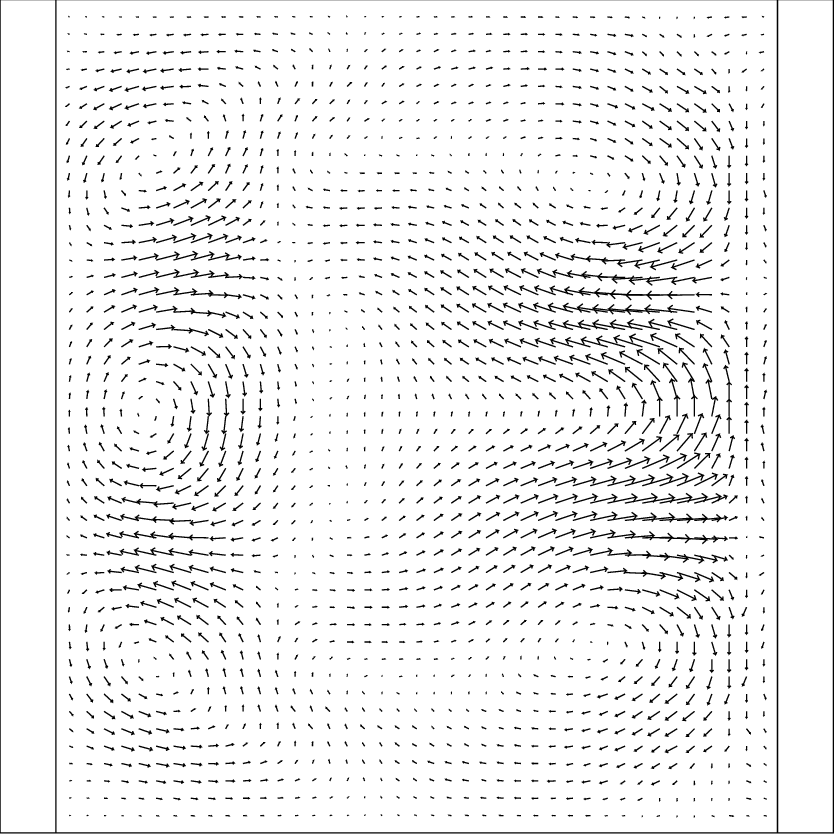}\hskip 2pt
\includegraphics[width=1.73in]{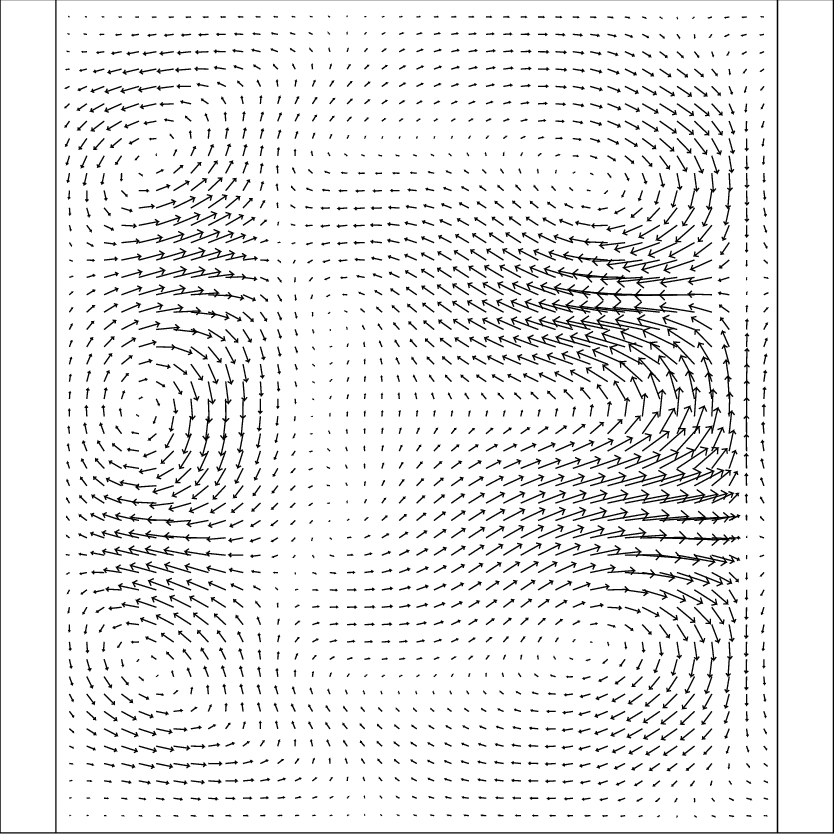}\hskip 2pt  \includegraphics[width=1.73in]{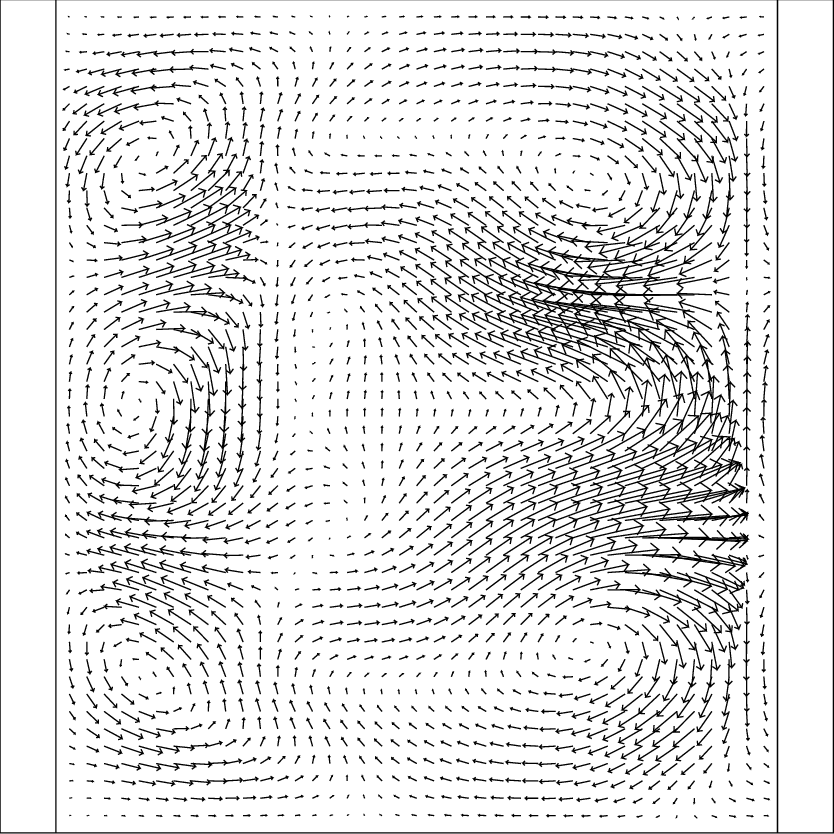} \\
\vskip 1ex
\includegraphics[width=1.73in]{vec-c-circular-square-Re0927-h096-average-t15008-1000000x-iz26.eps}\hskip 2pt
\includegraphics[width=1.73in]{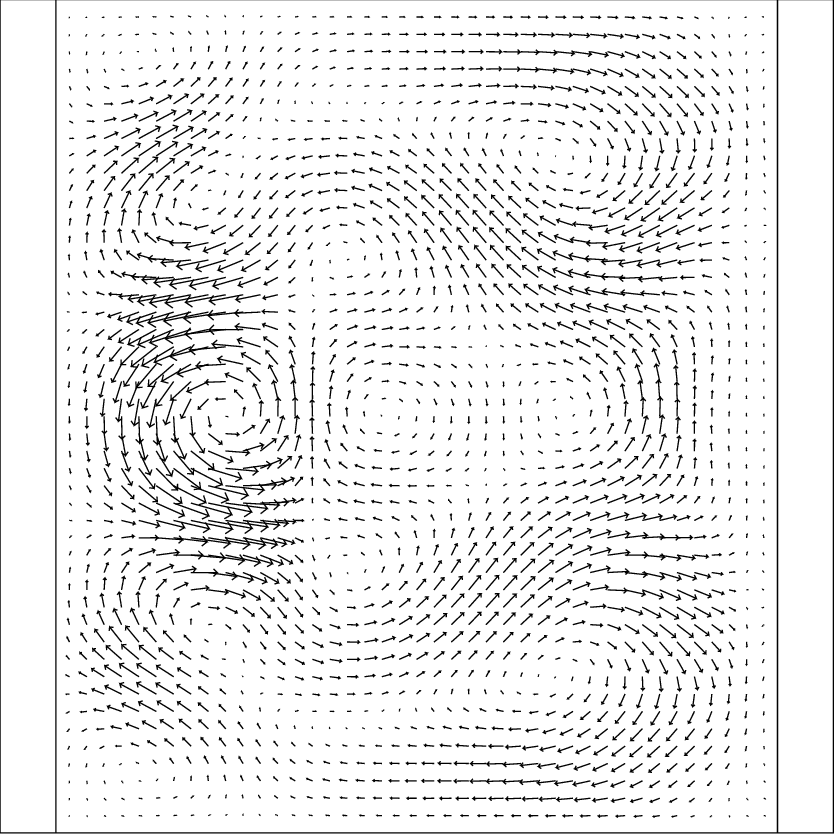}\hskip 2pt  \includegraphics[width=1.73in]{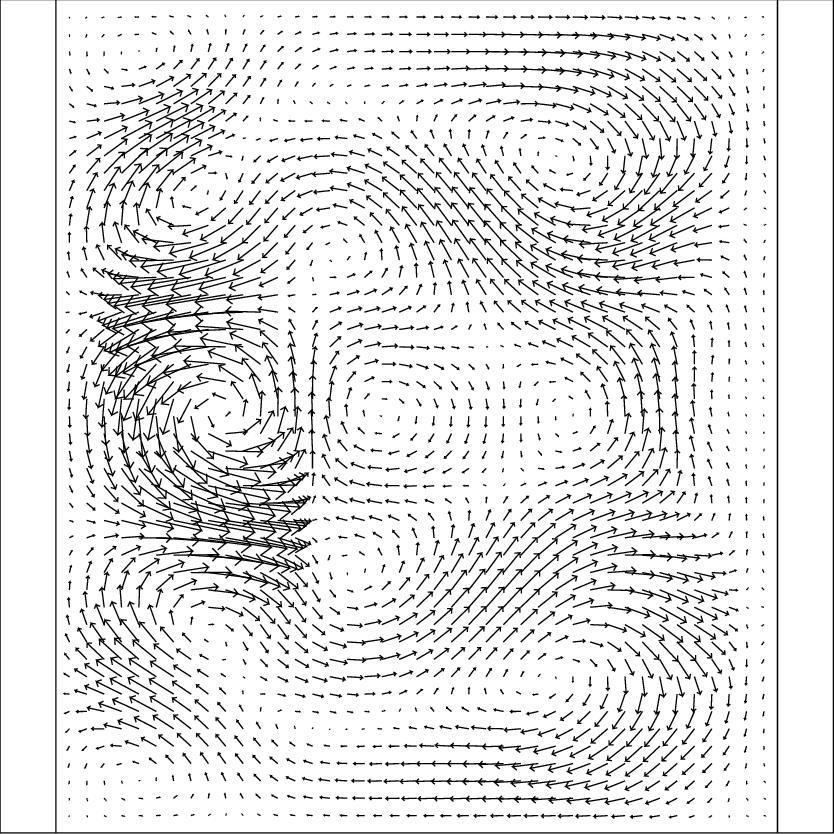} \\
\vskip 1ex
\includegraphics[width=1.73in]{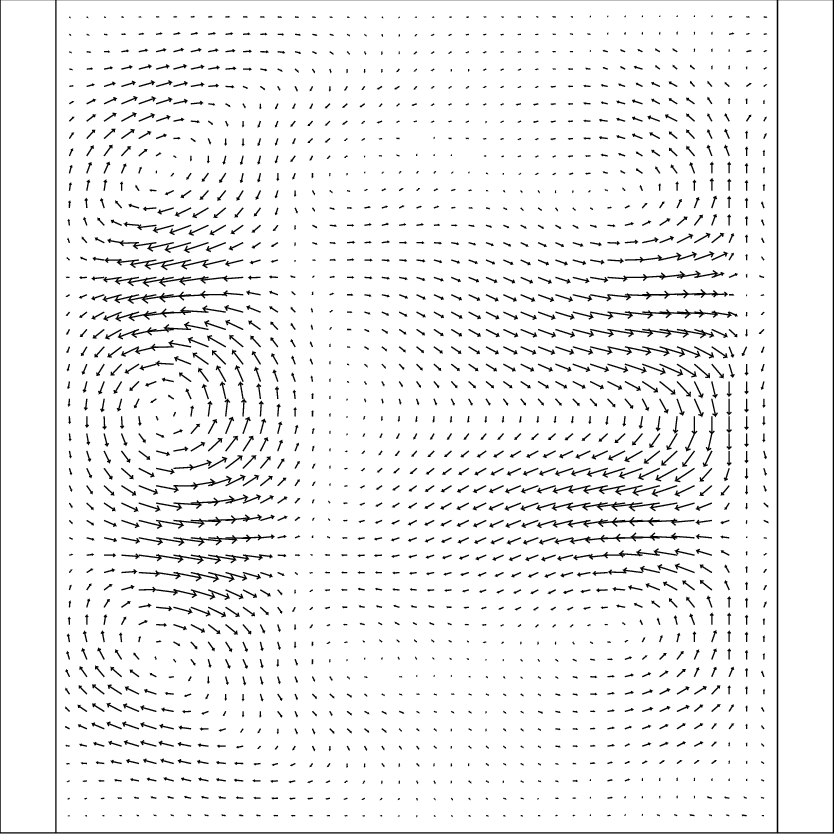}\hskip 2pt 
\includegraphics[width=1.73in]{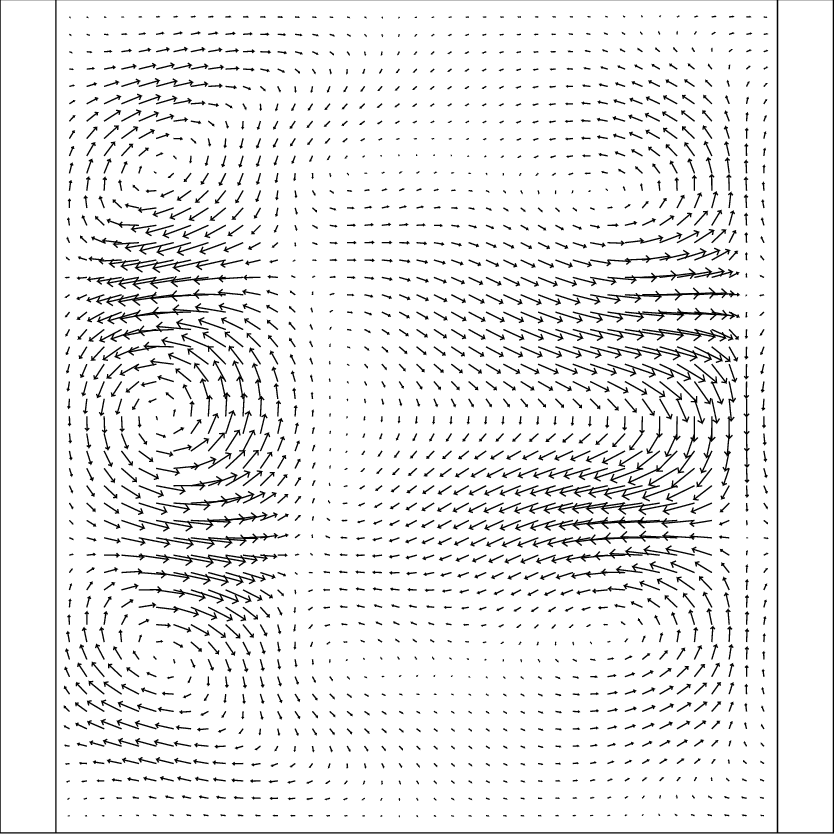}\hskip 2pt  \includegraphics[width=1.73in]{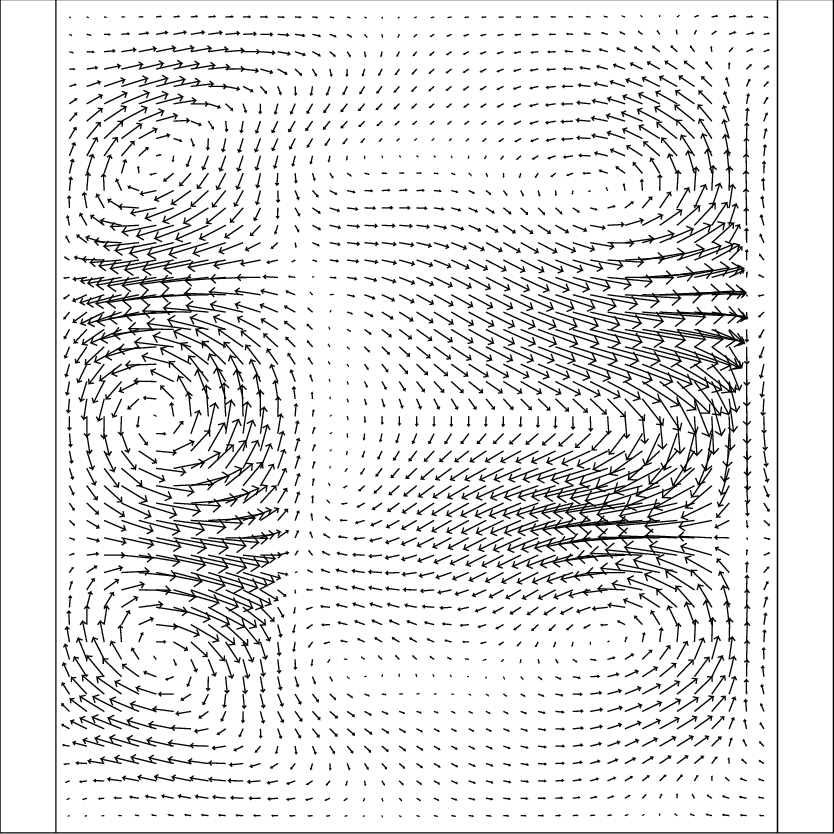} 
\end{center}
\caption{Oscillation velocity field (difference) projecting on  plane $x_3=0.25$  for Re=927 (left) at 
$t=$15006,  15008, and 15010 (from top to bottom),  Re=930 (middle) at $t=$14006.5, 14008.5, and
14010.5 (from top to bottom), and  Re=950 (right) at $t=$3007.65,  3009.65, and
 3011.65 (from top to bottom).  Velocity vectors have been magnified to enhance visibility.}\label{fig.14}
\end{figure}

\begin{figure} [!t]
\begin{center}
\leavevmode
\begin{minipage}{0.49\textwidth}
\phantom{1}\hskip 25pt \includegraphics[width=1.95in]{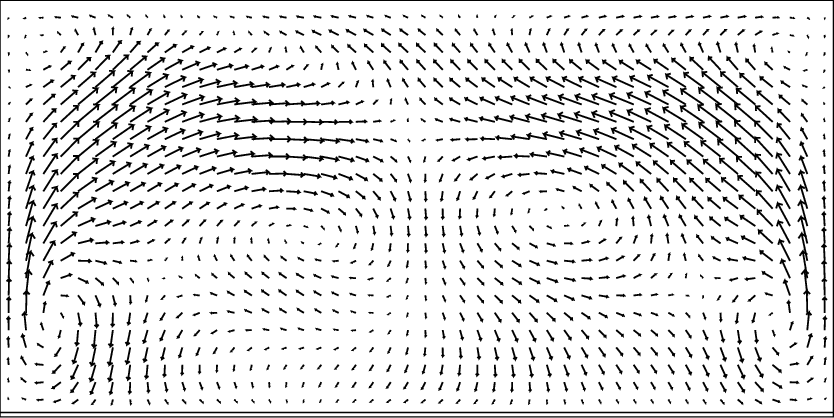} \\  
\phantom{1}\hskip 25pt \includegraphics[width=1.95in]{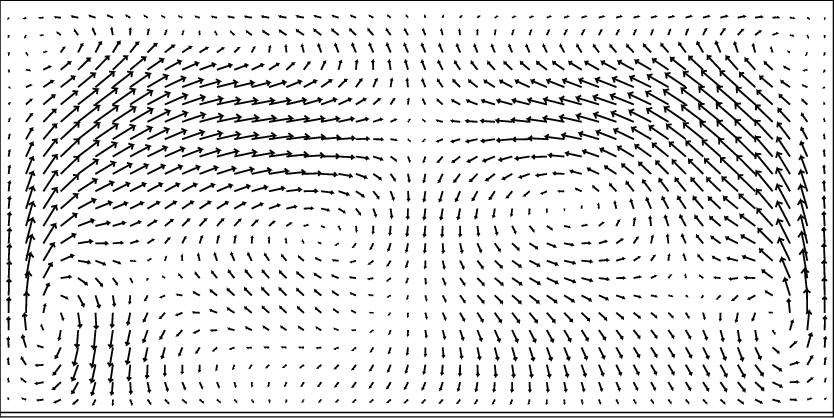} \\
\phantom{1}\hskip 25pt \includegraphics[width=1.95in]{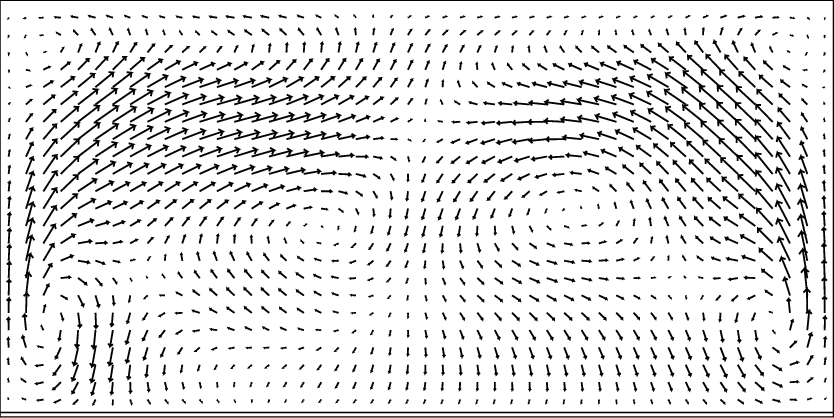} \\  
\phantom{1}\hskip 25pt \includegraphics[width=1.95in]{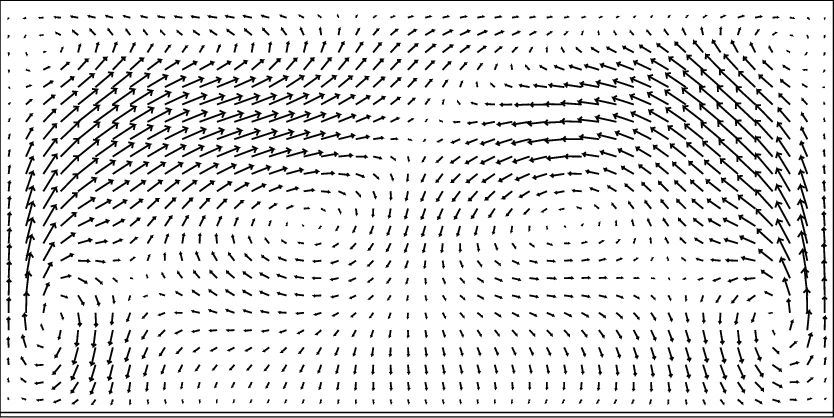} \\
\phantom{1}\hskip 25pt \includegraphics[width=1.95in]{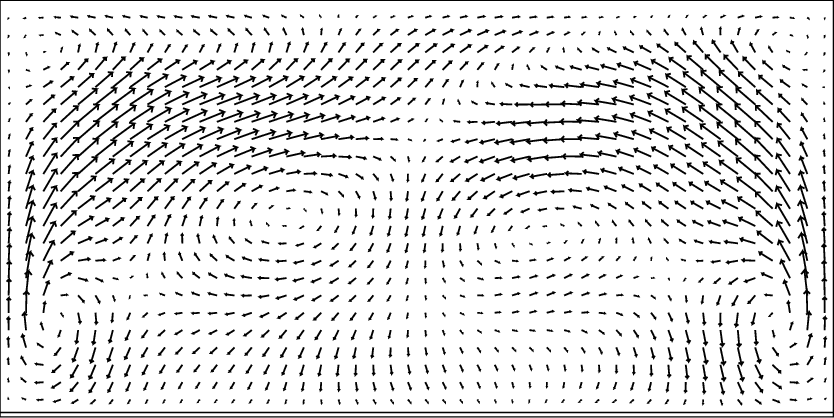} \\  
\hskip -5pt \includegraphics[width=0.38in]{cord-b3.eps}\
\includegraphics[width=1.95in]{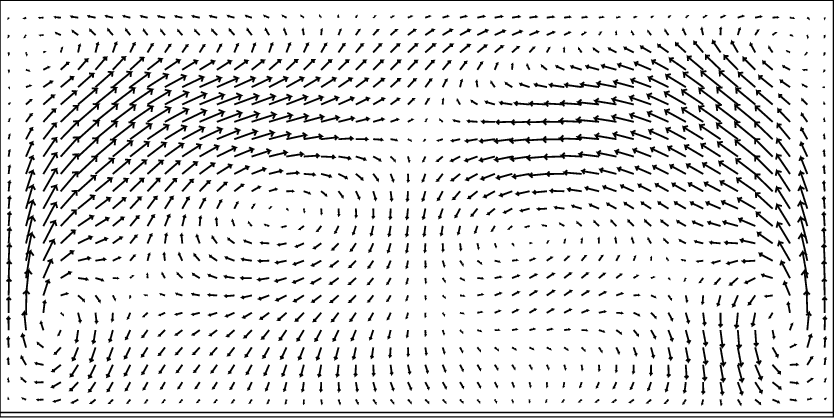} 
\end{minipage}
\hskip -5pt
\begin{minipage}{0.49\textwidth}
\phantom{1}\hskip 25pt \includegraphics[width=1.95in]{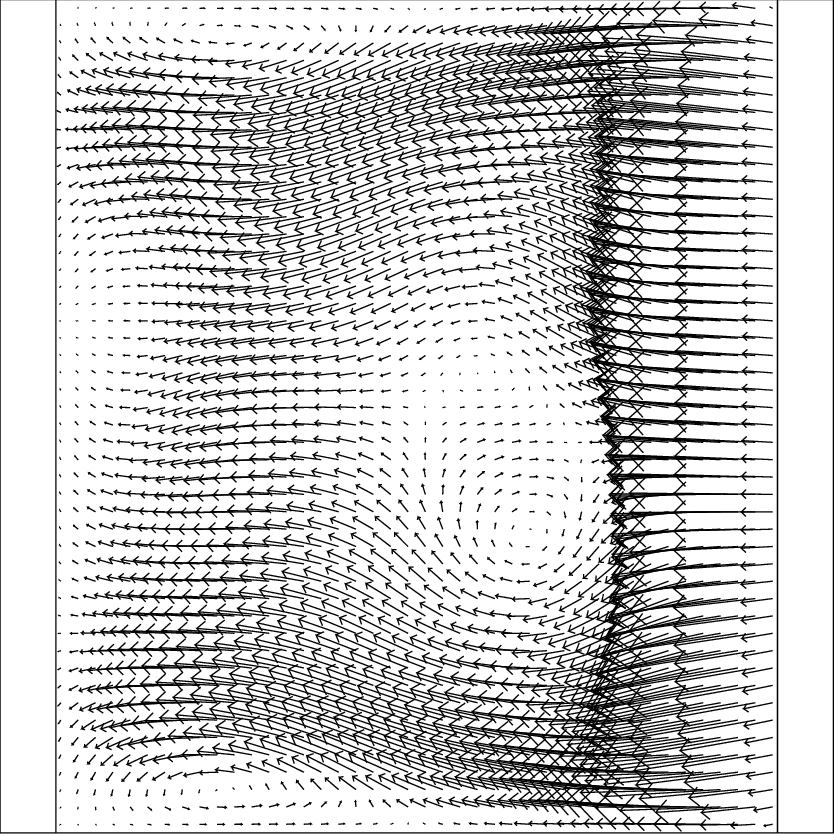} \\ 
\phantom{1}\hskip 25pt \includegraphics[width=1.95in]{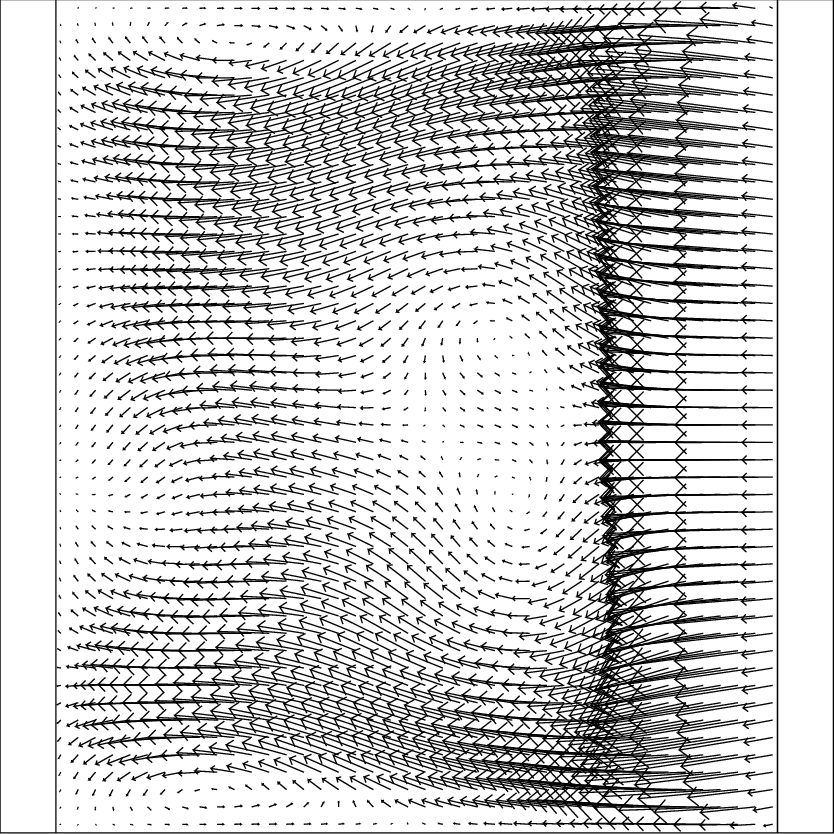} \\
\includegraphics[width=0.38in]{cord-c3.eps}\  \includegraphics[width=1.95in]{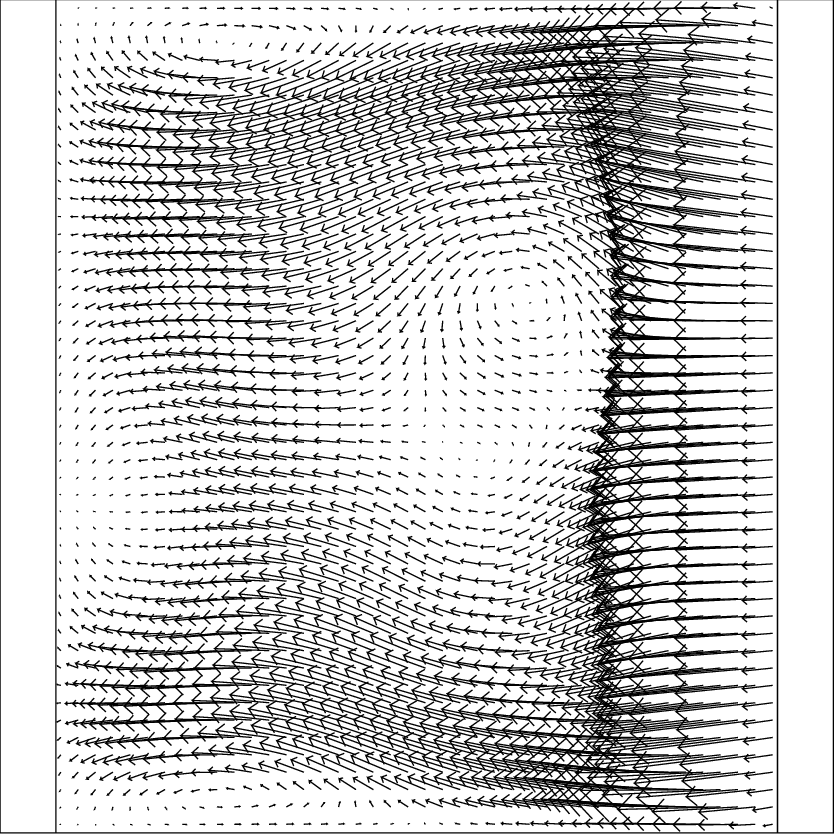} 
\end{minipage}
\end{center}
\caption{Velocity  field   projecting on the planes $x_1=55/96$ (left six) 
 at $t=3007.65$, 3008.65, 3009.65, 3010.65, 3011.65, and 3012.15 (from top to bottom) and $x_3=0.25$ (right three) 
 at $t=3007.65$, 3009.65, and 3011.65 (from top to bottom) for Re=950. }\label{fig.15}
\end{figure}

\begin{figure} [!t]
\begin{center}
\leavevmode
\includegraphics[width=4.75in]{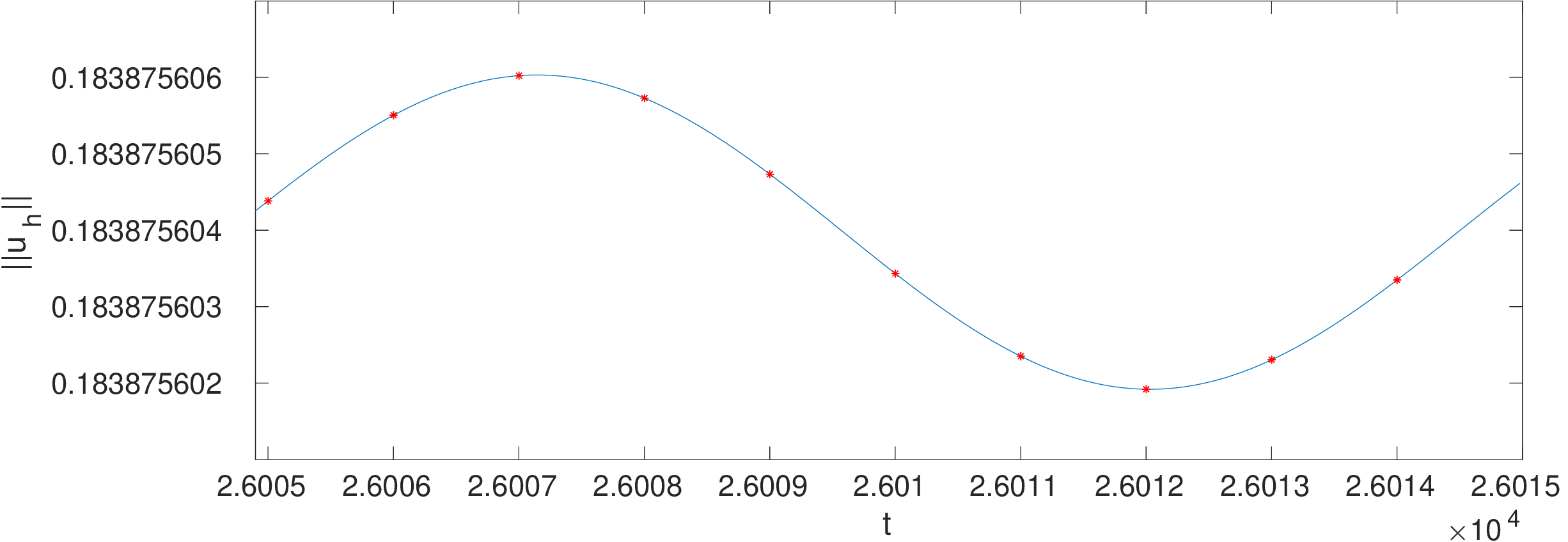}\\
\begin{minipage}{0.49\textwidth}
\includegraphics[width=0.38in]{cord-c3.eps}\ 
\includegraphics[width=2.1in] {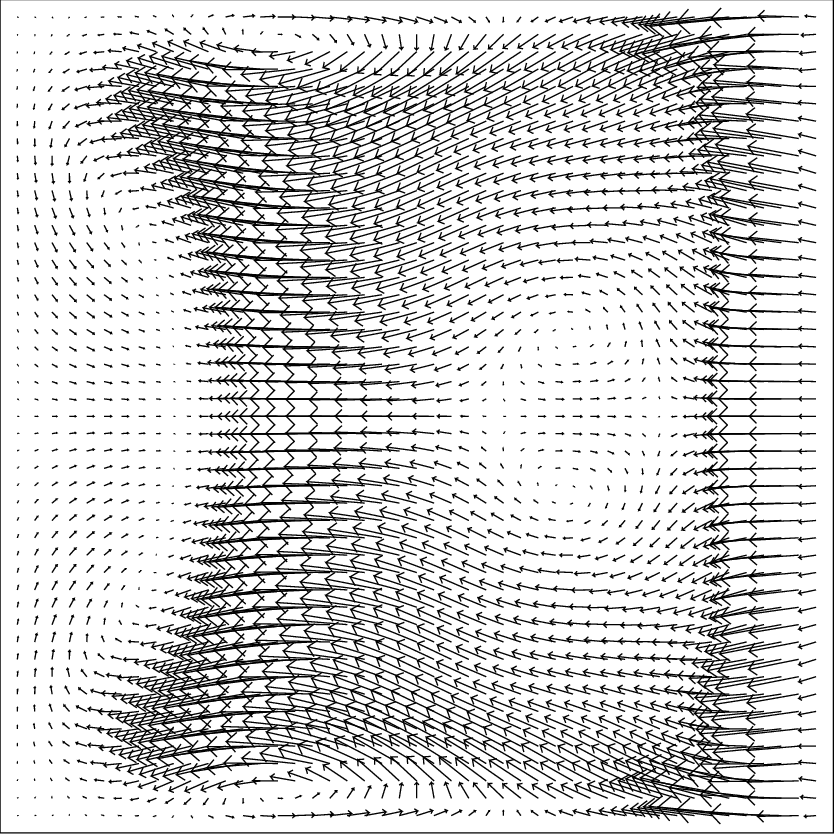}
\end{minipage}
\hskip -5pt
\begin{minipage}{0.49\textwidth}
\includegraphics[width=0.38in]{cord-a3.eps}\ 
\includegraphics[width=2.22in] {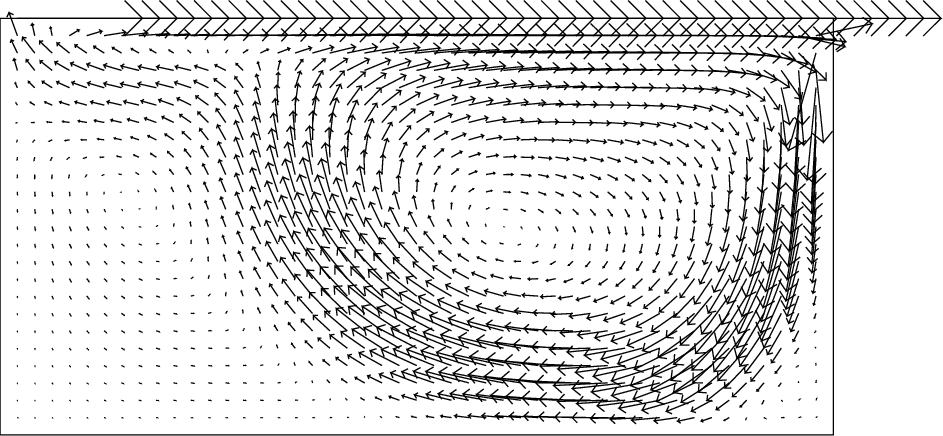}\\
\includegraphics[width=0.38in]{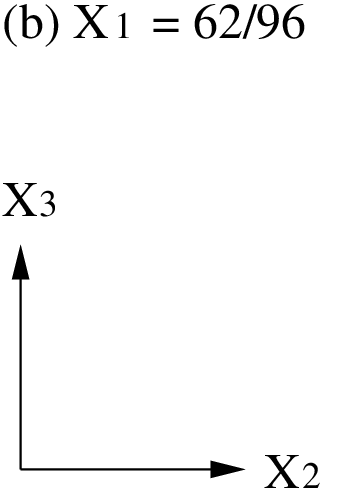}\hskip 5pt
\includegraphics[width=1.98in] {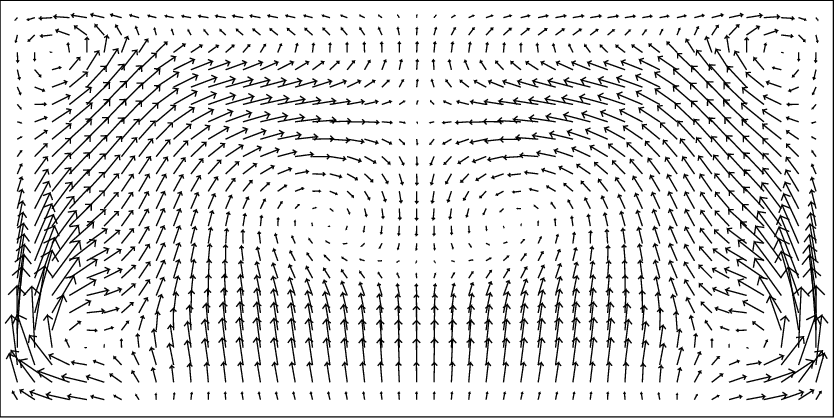} 
\end{minipage}
\end{center}
\caption{History of $\|\bu_h\|$ for one period in a shallow cavity with a unit square base for  $Re=1364$ (top), 
averaged velocity field projected on planes: $x_3=0.25$ (bottom left),  $x_2=0.5$ (middle right), 
and $x_1=62/96$ (bottom right) for  $h=1/96$ and   $\triangle t$=0.001. (In the bottom left and right 
plots, the vector scale is four times that of the actual one to enhance visibility.)}\label{fig.16}
\end{figure}

\begin{figure} [!t]
\begin{center}
\leavevmode
\includegraphics[width=1.88in]{vec-b-circular-square-Re0927-h096-average-t15006-2000000x-iy57.eps} \ 
\includegraphics[width=1.88in]{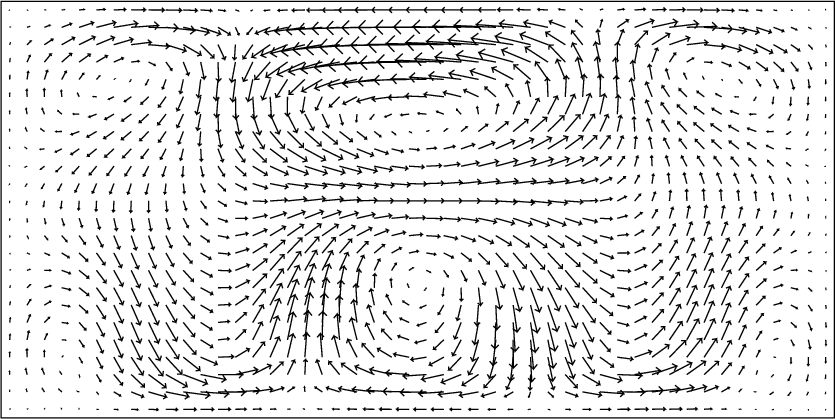} \\
\includegraphics[width=1.88in]{vec-b-circular-square-Re0927-h096-average-t15007-2000000x-iy57.eps} \  \includegraphics[width=1.88in]{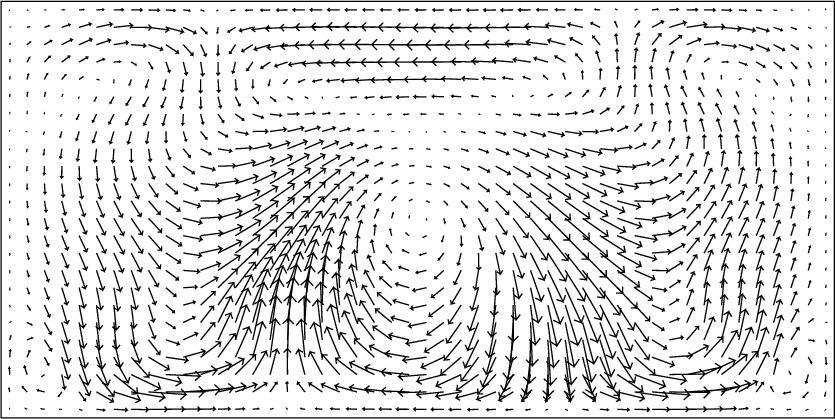} \\
\includegraphics[width=1.88in]{vec-b-circular-square-Re0927-h096-average-t15008-2000000x-iy57.eps} \  \includegraphics[width=1.88in]{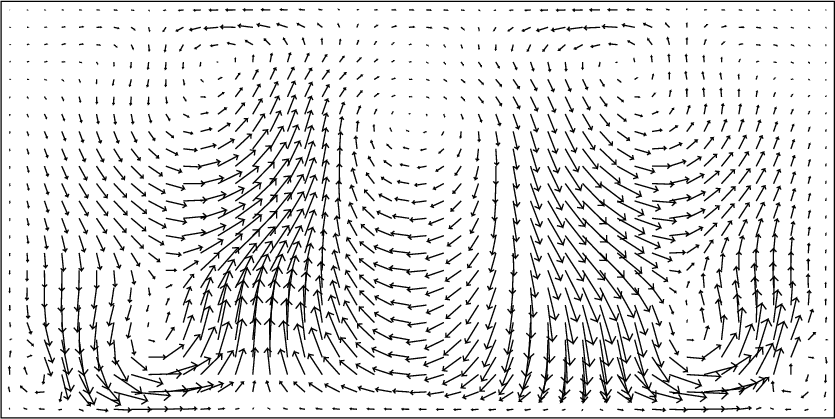} \\
\includegraphics[width=1.88in]{vec-b-circular-square-Re0927-h096-average-t15009-2000000x-iy57.eps} \  \includegraphics[width=1.88in]{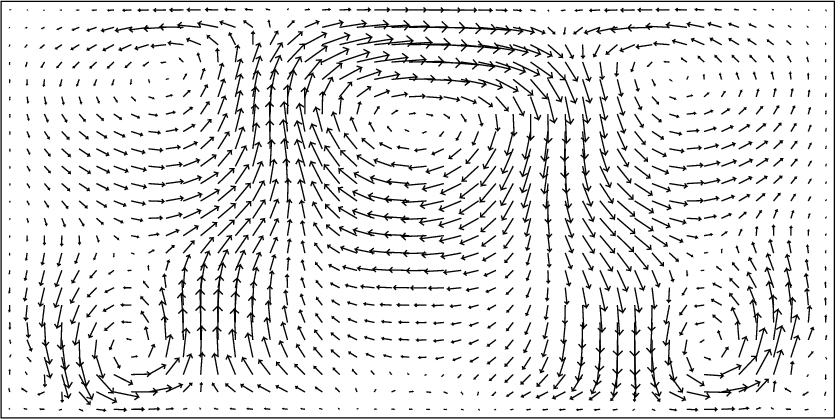} \\
\includegraphics[width=1.88in]{vec-b-circular-square-Re0927-h096-average-t15010-2000000x-iy57.eps} \  \includegraphics[width=1.88in]{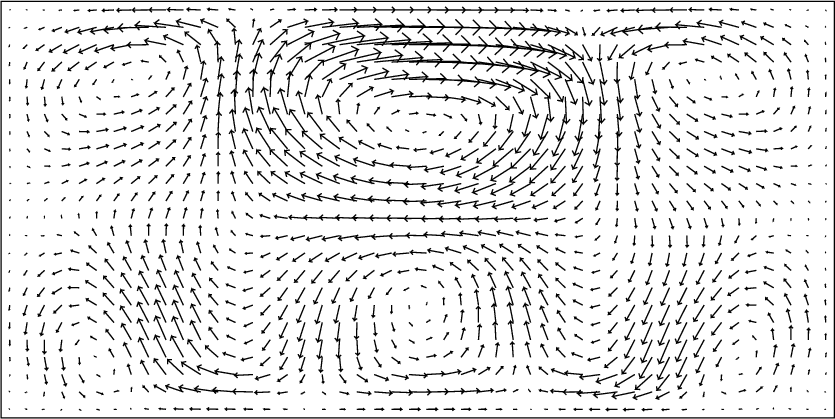} \\
\hskip -32pt \includegraphics[width=0.38in]{cord-b3.eps}\        \includegraphics[width=1.88in]{vec-b-circular-square-Re0927-h096-average-t15010d5-2000000x-iy57.eps} \  \includegraphics[width=1.88in]{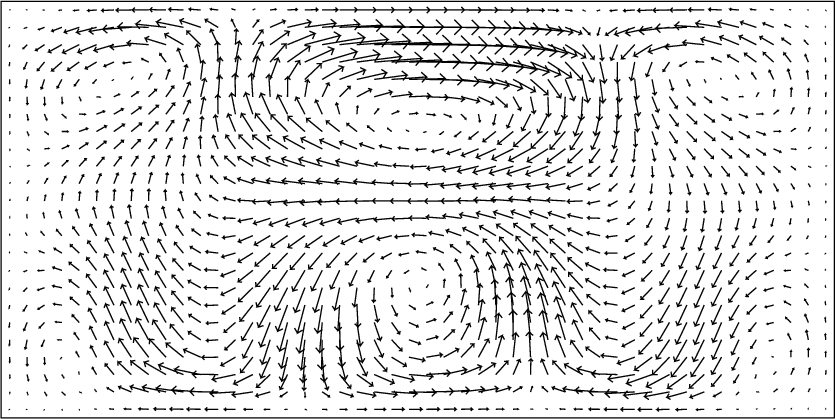} 
\end{center}
\caption{Oscillation mode projecting on  plane (i) $x_1=55/96$ for Re=927 in a 
semicircular cavity (left) at $t=$15006, 15007, 15008, 15009, 15010, and 15010.5 (from top to bottom) 
and (ii) $x_1=62/96$  for  Re=1364 (right) in a shallow cavity at $t=$ 26005,  26006.1, 26007.19, 
26008.29, 26009.39, and 26009.9 (from top to bottom).  Velocity vectors have been magnified to 
enhance visibility.}\label{fig.17}
\end{figure} 

\begin{figure} [!t]
\begin{center}
\leavevmode           
\includegraphics[width=1.91in]{vec-c-circular-square-Re0927-h096-average-t15006-1000000x-iz26.eps} \  \includegraphics[width=1.91in]{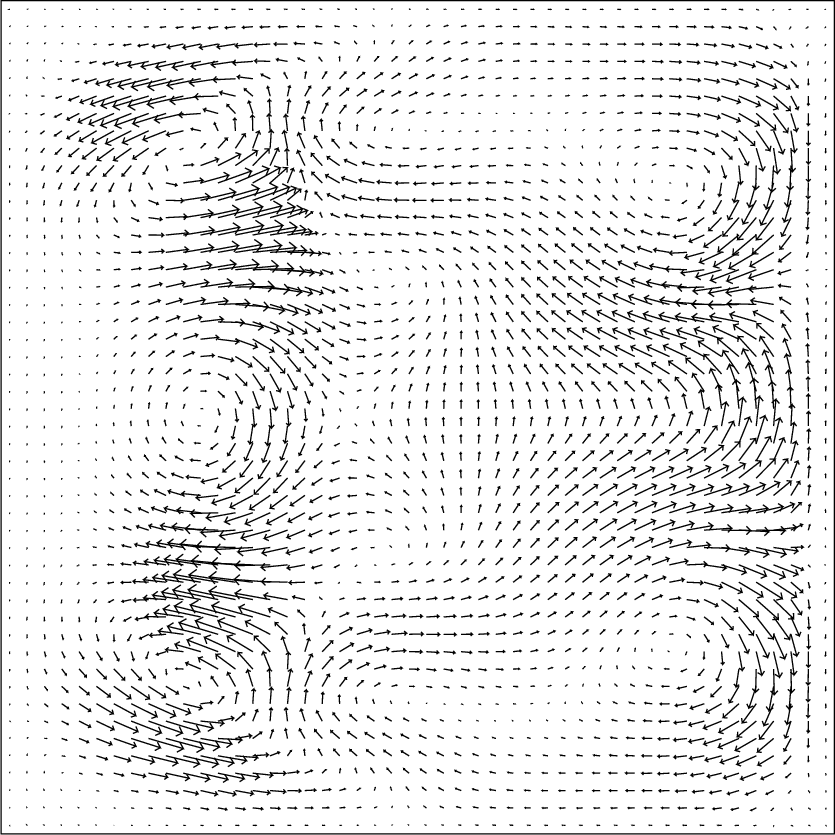} \\
\vskip 1ex
\includegraphics[width=1.91in]{vec-c-circular-square-Re0927-h096-average-t15008-1000000x-iz26.eps} \  \includegraphics[width=1.91in]{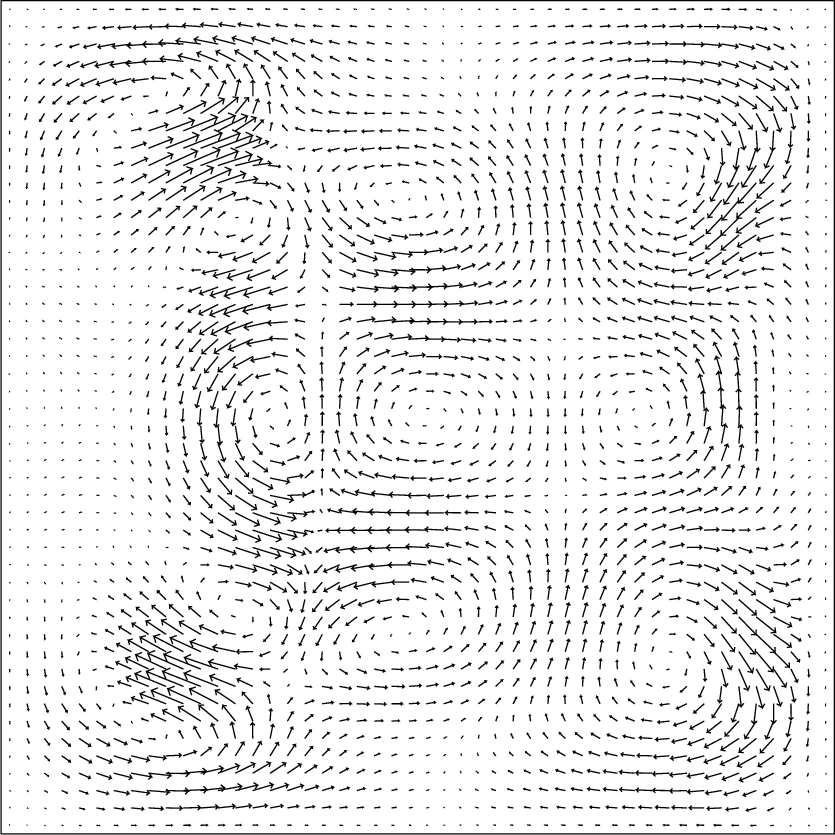} \\
\vskip 1ex
\hskip -32pt \includegraphics[width=0.38in]{cord-c3.eps}\ \includegraphics[width=1.91in]{vec-c-circular-square-Re0927-h096-average-t15010-1000000x-iz26.eps} \  \includegraphics[width=1.91in]{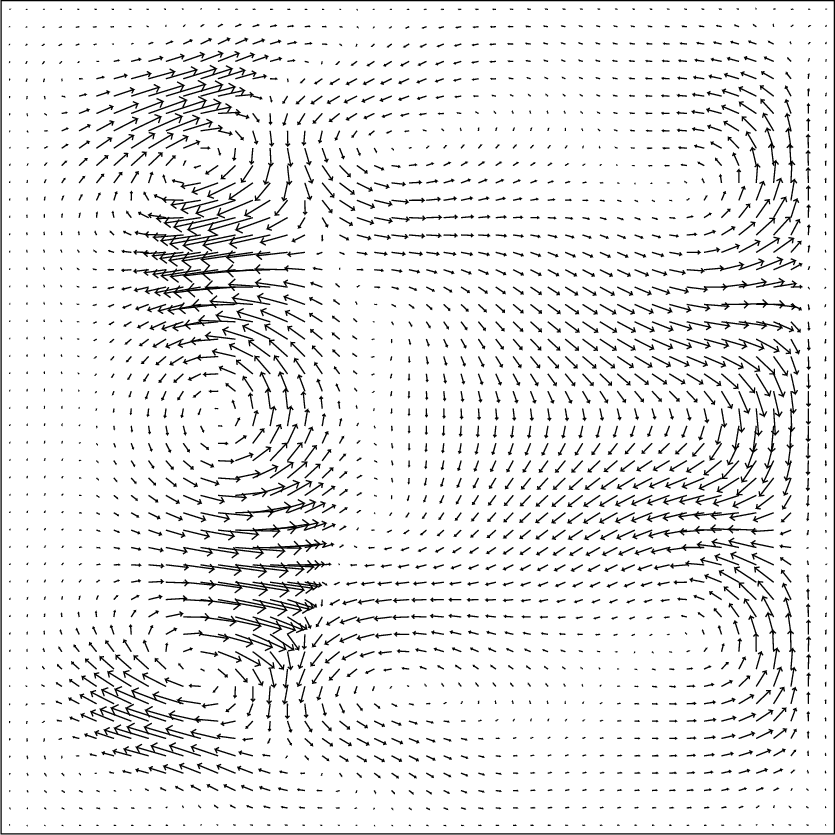} 
\end{center}
\caption{Oscillation velocity field (difference) projecting on  plane $x_3=0.25$  for Re=927 (left) at 
$t=$15006,  15008, and 15010 (from top to bottom) and for Re=1364 (right) in a shallow cavity
at $t=$ 26005,  26007.19, and 26009.39 (from top to bottom).  Velocity vectors have been magnified 
to enhance visibility.}\label{fig.18}
\end{figure}

Concerning the oscillation shown in the history of $L^2$-norm $\|\bu^n_h\|$, we have first computed the 
averaged velocity field from those obtained at different time instances denoted by ``*'' in one period as 
shown in the top plot of Figure  \ref{fig.10} for Re=927. Then we look into the  change of velocity field 
by comparing it to the average one obtained in the same period. The difference between 
averaged velocity field and the one at four time instances in the same time period are shown in 
Figure  \ref{fig.11} for Re=927.  The flow direction of the difference of velocity field  projected on 
plane $x_2=55/96$ at $t=15008$ (resp., $t=15010.5$) is opposed to the one at $t=15012.5$ (resp., 
$t=15015$), but the flow circulation pattern is almost the same. Similar behavior of flow direction 
and pattern are also found for the projection on plane $x_3=0.25$ in Figure   \ref{fig.11}.  
Since the difference between velocity field and the average one is a kind of ``oscillation mode'' (as 
shown in Figure  \ref{fig.11}), we shall look into the  difference of velocity field obtained from 
the first half of the period in following discussion. Even though the oscillating amplitude for Re=927 
decreases in time, the same oscillation mode has been obtained for Re=928 (whose oscillating amplitude 
increases in time as shown in Figure  \ref{fig.9}). For Re=930 and 950, both histories of $\|\bu_h\|$ do 
oscillate with  fixed amplitudes, respectively (see  Figures  \ref{fig.12a} and \ref{fig.12}). 
The angular frequencies are 0.7031317 and 0.6981317 for Re=930 and 950, respectively.
Their averaged velocity plots in  Figures  \ref{fig.12a} and \ref{fig.12}  show very  close similarity to those 
in  Figure  \ref{fig.10} for Re=927. 
To find out how oscillation mode evolves from Re=927 to 950,  we have compared the snapshots of oscillation mode  
for Re=930 and 950 at several instances  to those for Re=927 as shown in  Figures  \ref{fig.13}  and \ref{fig.14}.  
It is interesting to find out that the oscillation mode is almost the same for those three Reynolds numbers.
Those computational results for Re=927, 930, and 950 suggest that the oscillation mode 
in  Figures  \ref{fig.13} and \ref{fig.14} is the one associated with the Hopf bifurcation and originated at Re less than
the critical Reynolds number.

Since the oscillation amplitude of  $\|\bu^n_h\|$ in a time  period shown in 
Figure  \ref{fig.10} is about the order of $ 10^{-11}$ for Re=927, the change of velocity field from 
the average velocity field is very hardly to be observed. But for Re=950, the change of 
velocity field (due to the oscillation mode) can be observed in Figure \ref{fig.15} via the comparison to 
those in  Figure  \ref{fig.12}. 
On plane $x_1=55/96$  the size of vortices in the middle of those plots varies periodically and 
they oscillate slightly up and down. Similar on the plane $x_3=0.25$,  the size of vortices in the middle 
of those plots varies periodically. Another observation is that  the Taylor-G\"ortler-like vortices are not 
presented at the bottom of projected velocity field on the plane $x_1=55/96$ in Figures  \ref{fig.13}  and \ref{fig.15} 
for this case.

To find out the effect of semicircular shape on  lid-driven cavity flows, we have compared 
the above computational results at Re=927 to those in a shallow cavity with a unit square flat
bottom and height of 1/2. As reported in \cite{PanChiuGuoHe2023}, the critical Reynolds number 
for lid-driven flow in this shallow  cavity is between 1364 and 1365 for the mesh size $h=1/96$ 
and time step $\triangle t=0.001$.  The averaged velocity field and one period of $\|\bu_h\|$
for Re=1364 are shown in  Figure  \ref{fig.16}. The  averaged velocity field projected on 
$x_2=0.5$ is quite different from the one in a semicircular cavity shown in Figure \ref{fig.10}
due to the two different cross section shapes. But the similarity of averaged velocity field  on the planes 
$x_3=0.25$ and $x_1=62/96$ in  Figure  \ref{fig.16} can be found from those on planes $x_3=0.25$ and $x_1=55/96$
shown in Figure \ref{fig.10}. Snapshots of oscillation mode 
for Re=1364 from the first half of a period are presented in Figures \ref{fig.17} and  \ref{fig.18}. 
Comparing to those of Re=927 in a semicircular cavity (see Figures \ref{fig.17} and  \ref{fig.18}), we 
have found a very close similarity of the main flow circulation pattern. But in a shallow cavity, the main difference 
in Figure \ref{fig.17} is that   small vortices always exist at two lower corners, which
are triggered by the presence of two vertical side walls and downstream wall.

%
%

\section{Conclusion} 
In this article, we have studied numerically the transition from steady flow to oscillatory one in
a semicircular cavity of width and depth 1.   
Our simulation results show that the value of critical Reynolds number Re$_{cr}$  for the transition 
from steady flow to oscillatory  (a Hopf bifurcation) lie somewhere in the interval (927, 928)
for $h=1/96$.   The oscillating  angular frequency is between 0.70736  and 0.70345.  
The flow velocity oscillation at Re close to Re$_{cr}$  has been investigated in detail. 
We have visualized how the oscillating mode evolves for different Re values. Numerical results
indicate that the oscillation mode starts before Re=927 is the one associated with the Hope bifurcation 
at Re=930 and 950. Concerning the effect of semicircular shape on lid-driven cavity flow, we have found that 
the  oscillation mode in a semicircular cavity shows a close similarity to the one obtained in a shallow
cavity, but with some difference triggered by the presence of two vertical side walls and downstream wall
in a shallow cavity.


\end{document}